\documentclass[10pt,journal]{IEEEtran}
\IEEEoverridecommandlockouts
\usepackage{balance}
\usepackage{amssymb}
\usepackage{stmaryrd}
\usepackage{color}
\usepackage{multirow}
\usepackage{graphicx,times}
\usepackage{epstopdf}
\usepackage{indentfirst}
\usepackage{CJK}
\usepackage{amsmath}
\usepackage{amsfonts}
\usepackage{txfonts}
\usepackage{mathrsfs}
\usepackage{subfigure}
\usepackage{graphicx}
\usepackage{theorem}
\usepackage{url}
\usepackage{microtype}
\usepackage{fancyhdr}
\usepackage{bbm}
\usepackage{algorithm}
\usepackage{algorithmic}
\newcommand{\norm}[1]{\| #1 \|}
\def\BibTeX{{\rm B\kern-.05em{\sc i\kern-.025em b}\kern-.08em
		T\kern-.1667em\lower.7ex\hbox{E}\kern-.125emX}}

\begin{document}
	
\title{Over-The-Air Adversarial Attacks on Deep Learning Wi-Fi Fingerprinting}

\author{\IEEEauthorblockN{Fei Xiao, Yong Huang,~\IEEEmembership{Member,~IEEE}, Yingying Zuo, Wei Kuang,  Wei Wang,~\IEEEmembership{Senior Member,~IEEE}}
	\thanks{This work was supported by the Henan Province Key R\&D Program with Grant 221111210400. \textit{(Corresponding author: Yong Huang.)}}
	\thanks{Y. Huang is with the School of Cyber Science and Engineering, Zhengzhou University, Zhengzhou 450002, China (e-mail:yonghuang@zzu.edu.cn).}
	\thanks{F. Xiao is with Business School, Hubei University and School of Management, Huazhong University of Science and Technology, Wuhan 430074, China (e-mail: feixiao@hust.edu.cn).}
	\thanks{Y. Zuo, W. Kuang and W. Wang are with the School of Electronic Information and Communications, Huazhong University of Science and Technology, Wuhan 430074, China (e-mail:\{yingyingzuo, kuangwei, weiwangw\}@hust.edu.cn).}}

\maketitle

\begin{abstract}

Empowered by deep neural networks (DNNs), Wi-Fi fingerprinting has recently achieved astonishing localization performance to facilitate many security-critical applications in wireless networks, but it is inevitably exposed to adversarial attacks, where subtle perturbations can mislead DNNs to wrong predictions. Such vulnerability provides new security breaches to malicious devices for hampering wireless network security, such as malfunctioning geofencing or asset management. The prior adversarial attack on localization DNNs uses additive perturbations on channel state information (CSI) measurements, which is impractical in Wi-Fi transmissions. To transcend this limitation, this paper presents FooLoc, which fools Wi-Fi CSI fingerprinting DNNs over the realistic wireless channel between the attacker and the victim access point (AP). We observe that though uplink CSIs are unknown to the attacker, the accessible downlink CSIs could be their reasonable substitutes at the same spot. We thoroughly investigate the multiplicative and repetitive properties of over-the-air perturbations and devise an efficient optimization problem to generate imperceptible yet robust adversarial perturbations. We implement FooLoc using commercial Wi-Fi APs and Wireless Open-Access Research Platform (WARP) v3 boards in offline and online experiments, respectively. The experimental results show that FooLoc achieves overall attack success rates of about 70\% in targeted attacks and of above 90\% in untargeted attacks with small perturbation-to-signal ratios of about -18~dB.

\end{abstract}


\section{Introduction}

In wireless networks, accurate device location information is increasingly desired to support many security-critical applications, such as device authentication and access control~\cite{liu2019wireless, liu2019collaborative}. To achieve this, Wi-Fi fingerprint based indoor localization recently has gained astonishing performance via benefiting from the advances in deep neural networks (DNNs)~\cite{wang2016csi,nowicki2017low,abbas2019wideep,wang2020indoor}, which, however, are shown to be susceptible to adversarial attacks~\cite{szegedy2013intriguing,goodfellow2014explaining, eykholt2018robust}. In such attacks, minimal perturbations on genuine input samples can steer DNNs catastrophically away from true predictions. By exploiting these vulnerabilities, malicious devices have the potential to manipulate their localization results and cause the breakdown of wireless geofencing~\cite{sheth2009geo,pan2008digital}, asset management, and so on. Thus, it is of great importance to investigate the extent to which DNN powered indoor localization is vulnerable to adversarial attacks in the real world. 

Despite the great importance, no existing study explores over-the-air adversarial attacks on indoor localization DNNs in the physical world. The prior work~\cite{patil2021adversarial} investigates adversarial attacks on indoor localization DNNs and simply adds perturbation signals to original signals likewise generating adversarial images in the computer vision domain. However, additive perturbations can not characterize the impact of Wi-Fi training signals on CSI measurements, thus rendering them infeasible in over-the-air attacks. Moreover, these approaches~\cite{tung2016analog,fang2014you} trigger attacks by directly converting genuine CSI fingerprints into targeted ones, which are suitable for attacking single-antenna APs. Yet, they are physically unrealizable in widely-used multi-antenna Wi-Fi systems due to the one-to-many relationship between transmitting and receiving signals. In addition, this study~\cite{cominelli2021ieee} proposes a CSI randomization approach to distort device location information. Though this approach can trigger untargeted adversarial attacks, it lacks the capability of misleading location predictions close to chosen spots, i.e., targeted attacks. In addition, the random perturbations are not smooth and will cause significant disturbance in the original signals, rendering them easy to be detected. Thus, no existing work is suitable for launching adversarial attacks on Wi-Fi fingerprinting DNNs in the real world.

\begin{figure}
	\centering
	\includegraphics[width=1\linewidth]{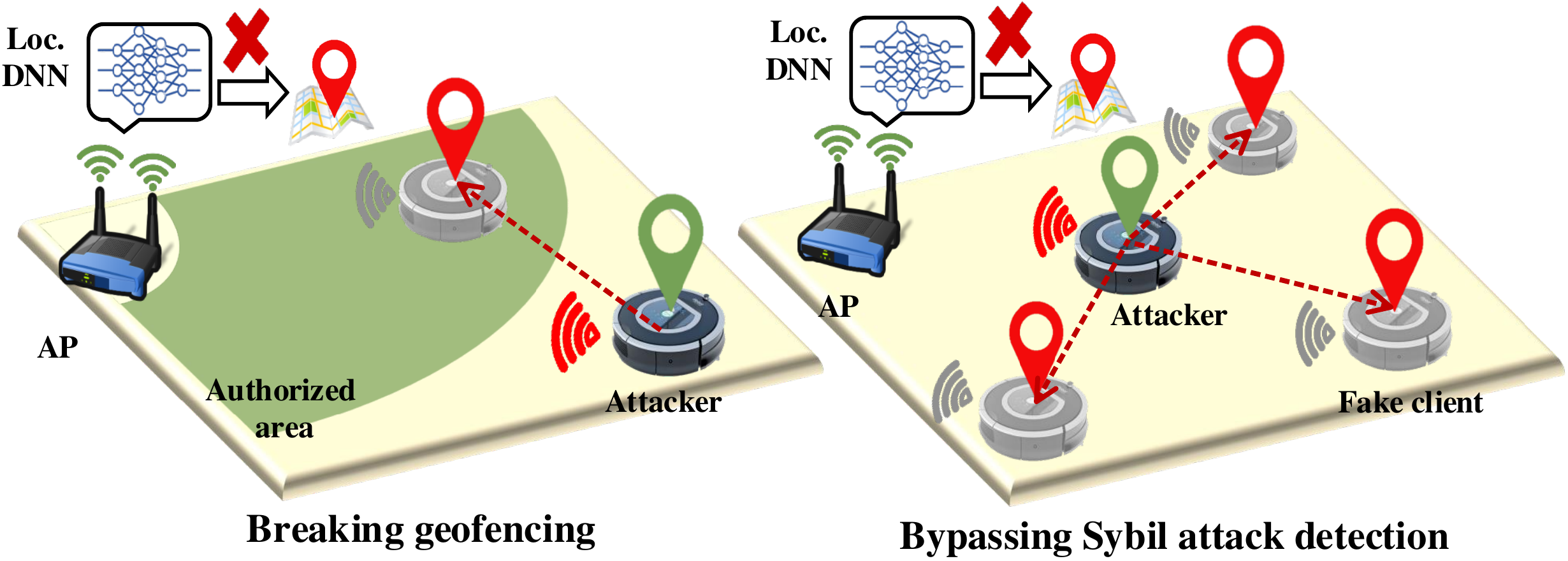}
	\caption{Attack cases with over-the-air adversarial attacks.}
	\label{fig:attackscenario}
\end{figure}

In this paper, we investigate a new type of adversarial attack that deceives indoor localization DNNs over realistic wireless channels. In particular, our attack model includes a Wi-Fi AP and an attacker. The AP holds a well-trained DNN for indoor localization using uplink CSI signatures as inputs. The attacker, i.e., a malicious client device, manipulates its Wi-Fi training signals and transmits them to the AP over the air, with the purpose of fooling the localization DNN. In this way, the AP receives the falsified signals from the attacker, generates perturbed uplink CSI signatures, and feeds them into the DNN for device localization. As demonstrated in Fig.~\ref{fig:attackscenario}, over-the-air attacks can rise severe security issues in wireless networks. An outside attacker can be empowered to break the geofencing of a Wi-Fi AP by camouflaging itself within authorized areas to gain wireless connectivity. Moreover, an attacker can bypass Sybil attack detection to deplete valuable bandwidth by pretending multiple fake clients at the same location~\cite{newsome2004the,huang2021detecting}.

We argue that the major obstacle to realizing such over-the-air adversarial attacks is that the uplink CSI estimated at the victim AP is unknown to the attacker and thus effective channel perturbations cannot be generated before each attack. To tackle this problem, we observe that the similarity between uplink and downlink CSIs can be exploited for launching adversarial attacks over the air. In Wi-Fi networks, downlink CSIs can be easily obtained from the AP's broadcasting packets, such as beacon frames. When one attacker stays at one spot, its uplink and downlink transmissions would experience similar multipath propagations and thus have similar CSI fingerprints~\cite{tse2005fundamentals}. Hence, the attacker can take benefits of accessible and informative downlink CSIs to generate adversarial perturbations locally without knowing the exact uplink CSIs that are fed into localization DNNs by the AP.

Toward this end, we present FooLoc, a novel system that fools localization DNNs by launching over-the-air adversarial attacks. Specifically, before each attack, FooLoc takes obtainable downlink CSIs as a reasonable substitute of the corresponding uplink ones and trains an adversarial perturbation locally. Then, it applies the well-trained perturbation on its own transmitted signals for manipulating the corresponding uplink CSI signatures received by the AP. In this way, FooLoc is capable of deceiving the localization DNN to output desired yet wrong location estimates over real wireless channels.

To realize the above idea, we address the following two challenges.

\textit{1) How to design realizable adversarial perturbations that are suitable for Wi-Fi transmissions?} Most adversarial attacks are based on additive perturbations and require the ability to individually alter each element of an input sample, which, however, is physically unrealizable for over-the-air perturbations. Specifically, in Wi-Fi communications, a physical layer training symbol has a multiplicative relationship with a channel response in the frequency domain~\cite{tse2005fundamentals}, thus rendering additive perturbations on Wi-Fi CSIs infeasible. Moreover, for a multi-antenna receiver, one training symbol of each subcarrier corresponds to multiple received symbols during channel estimation, implying a one-to-many relationship between the elements of one perturbation and one CSI measurement. Based on the discovered multiplicative and repetitive properties, we formulate the novel over-the-air perturbations on uplink CSIs and further derive the adversarial perturbations for targeted and untargeted attacks on indoor localization DNNs.

\textit{2) How to efficiently craft imperceptible yet robust adversarial perturbations under environmental noise?} Due to the random nature of environmental noise, two CSI measurements from the same spot are unlikely to be exactly the same. Consequently, one perturbation that is generated for one specific CSI may not generalize well to another one. To tackle this challenge, we propose a generalized objective function integrating both targeted and untargeted attacks and reasonably formulate the adversarial perturbation generation as a box-constrained optimization problem. In this optimization problem, we ensure the robustness of adversarial perturbations by seeking a universal perturbation that works well on all CSI measurements from the same spot and guarantee their imperceptibility by maximizing the perturbation smoothness and limiting the perturbation strength at the same time. Moreover, to ease the difficulty of problem optimization, we further transform the constrained problem into an equivalent unconstrained one.

\textbf{Summary of Results.} We implement FooLoc using commercial Wi-Fi APs for offline experiments and Wireless Open-Access Research Platform (WARP) v3 boards~\cite{anand2010warplab} for online experiments. In offline experiments, FooLoc obtains attack success rates (ASRs) of 73.0\% and 93.4\% for targeted and untargeted attacks, respectively, on average. In online experiments, FooLoc achieves mean ASRs of 71.6\% and 99.5\% for targeted and untargeted attacks, respectively. Moreover, FooLoc has small perturbation-to-signal ratios (PSRs) of about -18~dB in two settings.

\textbf{Contributions.} The main contributions of this work are summarized as follows. 
\begin{itemize}
	\item We propose FooLoc, which exploits the similarity between uplink and downlink CSIs to launch over-the-air adversarial attacks on Wi-Fi localization DNNs.
	\item We discover the multiplicative and repetitive impacts of over-the-air perturbations on CSI fingerprints in Wi-Fi localization systems.
	\item We propose an efficient algorithm to generate imperceptible and robust adversarial perturbations against localization DNNs over realistic Wi-Fi channels.
	\item We implement FooLoc on both commercial Wi-Fi APs and WARP wireless platforms, respectively, to demonstrate its effectiveness in different environments.
\end{itemize}

\section{Attack Model and Wi-Fi CSI Signatures}

\subsection{Adversarial Attacks on Indoor Localization}

In this paper, we consider a general Wi-Fi network, where one fixed AP with multiple antennas provides wireless connectivity for many single-antenna clients, such as smartphones and vacuum robots. The AP has the capability of device localization for delivering location based services, such as user monitoring and access control. Moreover, we focus on deep learning (DL) based indoor localization systems, which exploit accessible and fine-grained Wi-Fi CSIs as location fingerprints. Considering the randomness of CSI phases, most fingerprinting systems rely on CSI amplitudes~\cite{wang2016csi,wang2015deepfi}. Hence, such DL models are assumed to accept CSI amplitudes as input features and output 2D continuous-valued location estimations.

To fool such localization systems in reality, we consider the over-the-air adversarial attacks by exploiting the vulnerabilities of DNNs~\cite{szegedy2013intriguing}. In this scenario, a malicious attacker, as a client device, can not directly manipulate the input values of DL models used by the AP. Instead, it can attack a DL model only via modifying its own transmitted Wi-Fi signals. In this paper, we mainly consider white-box DL models, of which the attacker knows their exact structures as well as trained parameters. For black-box models that are unknown to the attacker, we will discuss the feasibility of triggering adversarial attacks on them in our offline experiment. Furthermore, the attacker has no access to uplink CSI measurements that are used for model training and testing. Yet, it has the ability to move in the targeted area and collects corresponding downlink CSI measurements. For example, the attacker could be a vacuum robot, which moves between different spots to automatically collect Wi-Fi CSI fingerprints~\cite{sen2012you,ayyalasomayajula2020deep}. 

In addition, we assume that the attacker knows its own location information when launching adversarial attacks for misleading location based services provided by the AP. Moreover, we consider targeted and untargeted adversarial attacks on localization DNNs. Specifically, in targeted attacks, the attacker aims to force the localization model to output a location estimate that is as close as possible to a chosen spot. When comes to untargeted attacks, it only wants to be localized far away from its true location. 

Such over-the-air adversarial attacks can be exploited to deceive localization DNNs~\cite{wang2016csi,wang2015deepfi} for hampering security of wireless networks. The example attack scenarios include 1) breaking geofencing: a Wi-Fi AP holds a device localization model and provides wireless connectivity only to clients that are within a certain area. In this scenario, an attacker stays outside of the area and can trigger over-the-air adversarial attacks to camouflage itself inside authorized areas for gaining wireless connectivity; 2) bypassing Sybil attacker detection: a Wi-Fi AP uses a localization model to detect potential Sybil attackers based on their locations. Using over-the-air adversarial attacks, an attacker can masquerade many fictitious clients that are seemingly from different locations to deplete valuable bandwidth at a low cost.

\begin{figure}
	\centering
	\includegraphics[width=1\linewidth]{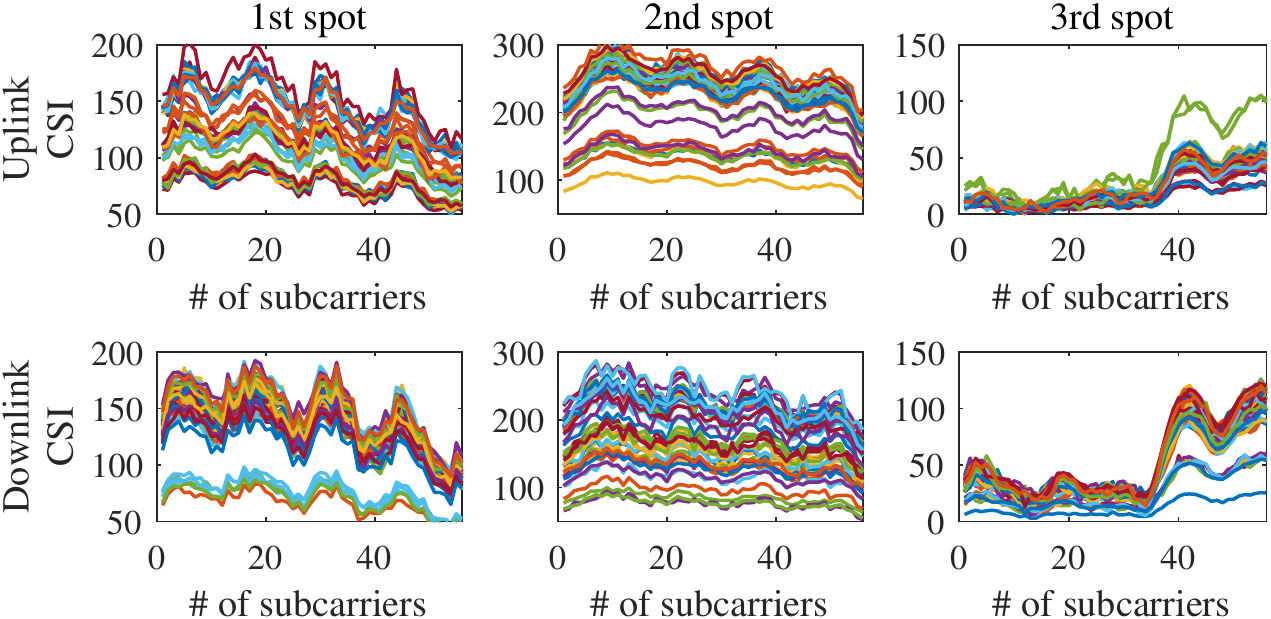}
	\caption{Uplink and downlink CSI measurements at different spots. The distances of 1st spot to 2nd and 3rd spots are 0.3~m and 1.2~m, respectively.}
	\label{fig:feasibilitystudy}
\end{figure}

\subsection{Wi-Fi CSI Fingerprints}

Basically, channel state information characterizes the signal propagation among a pair of Wi-Fi transceivers in a certain environment. The IEEE 802.11n/ac/ax Wi-Fi protocols divide a Wi-Fi channel into $ K $ orthogonal subcarriers and assign $ K $ pre-defined long training field signals (LTFs) for them. For the $ k $-th subcarrier, the transmitter sends a training signal $ s_k $, and accordingly the receiver obtains a signal $ y_k $. With the knowledge of $ s_k $, the receiver can estimate the current channel response $ h_k $ between them as
\begin{align} \label{eq: csi estimation}
	h_k  = y_k/s_k.
\end{align}
Due to multipath effects, each channel response $ h_k $ can be further modeled as the composition of one direct path and multiple reflected ones~\cite{tse2005fundamentals}, which can be formulated as
\begin{align} \label{eq: multipath}
	h_{k} =  \alpha_0 e^{j2\pi \tau_0 f_k} +  \sum_{l}\alpha_l e^{j2\pi \tau_l f_k}   + n_k,  
\end{align}
where $ n_k $ is the complex Gaussian noise. Moreover, $ \alpha_0 $ and $ \tau_0 $ represent the signal propagation attenuation and time delay of the direct path, respectively, and $ \alpha_l $ and $ \tau_l $ are those of the $ l $-th reflected path. From the above equation, we can see that Wi-Fi CSI measurements are highly dependent on transceiver locations as well as environmental reflectors. For a fixed-position AP-client pair, uplink and downlink signals would travel through the alike line-of-sight distances as well as similar incident-reflecting paths. The above geometric properties together contribute to nearly-identical path loss and time delay, thus resulting in similar channel responses. Such similarity enables an adversary to replace unknown uplink CSIs with the corresponding downlink ones for generating adversarial perturbations.  

We conduct some preliminary studies to verify the similarity between paired uplink and downlink CSI fingerprints. To do this, we use two off-the-shelf Wi-Fi APs with Atheros CSI Tool~\cite{xie2015precise} to record CSI measurements of 56 subcarriers. In our experiments, we fix one AP at a certain location and place the other at three different spots. As plotted in Fig.~\ref{fig:feasibilitystudy}, we can observe that similar change patterns are shared in uplink and downlink measurements corresponding to the same spot. This is because when the locations of two APs are fixed, the uplink and downlink signals would experience similar multipath propagations as indicated in Eq.~\eqref{eq: multipath}. It is worth noting that the occurrence of multiple clusters of CSI measurements in each subfigure is caused by automatic gain control on the receiver side for maintaining a suitable power level. In addition, it also can be found that the similarity in CSI measurements increases as the distance between two spots decreases. The above observations verify that uplink and downlink CSI measurements are highly similar, providing an exciting opportunity to launch over-the-air adversarial attacks on DL indoor localization systems.

\section{Over-The-Air Adversarial Attacks}\label{sec:system design}

\begin{figure}
	\centering
	\includegraphics[width=1\linewidth]{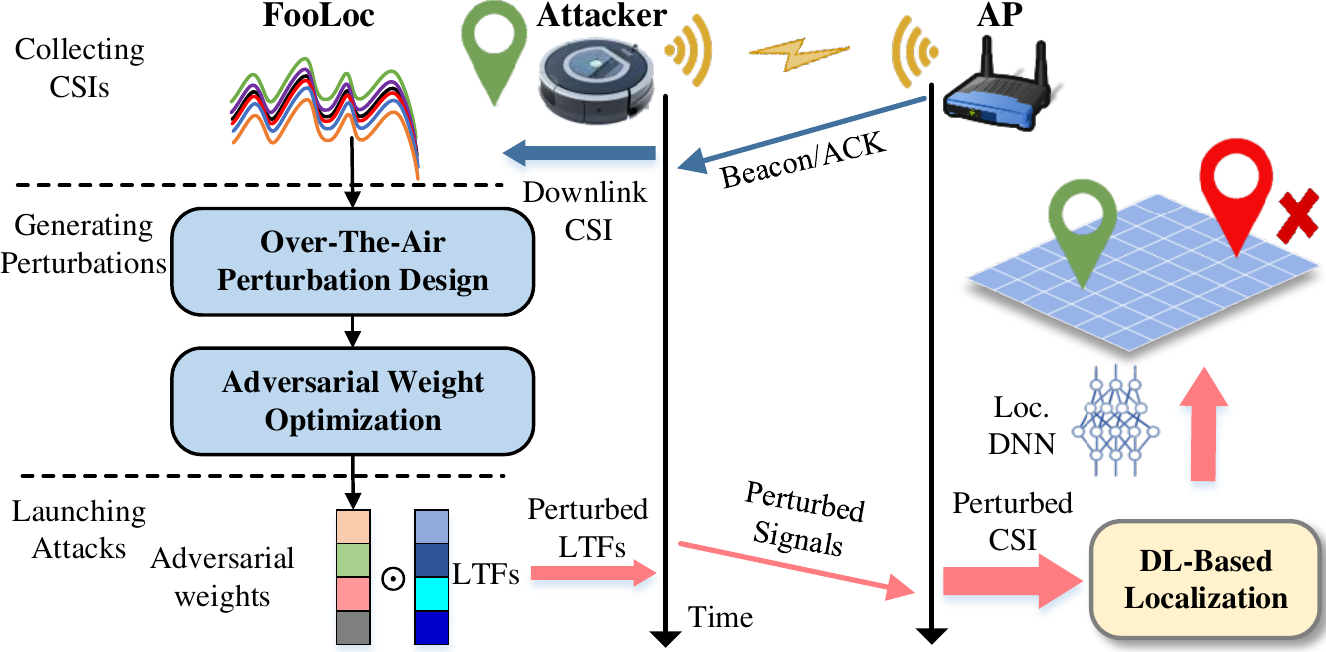}
	\caption{Workflow of FooLoc for launching over-the-air adversarial attacks on DL indoor localization.}
	\label{fig:systemoverview}
\end{figure}

\subsection{Overview of FooLoc}

FooLoc is a novel system that fools Wi-Fi CSI fingerprinting localization DNNs via launching over-the-air adversarial attacks. 
As depicted in Fig.~\ref{fig:systemoverview}, FooLoc runs on the attacker and helps it to spoof the localization DNN used by the AP. Specifically, before each attack, the attacker first stays at one spot and receives downlink packets, such as beacon and acknowledgment (ACK) frames~\cite{xiao2005ieee}, from the targeted AP.
Then, FooLoc generates a set of well-crafted adversarial weights based on its knowledge of the victim model.
After that, it multiplies the adversarial weights with genuine LTFs and sends their product results to the AP over the air.
Once receiving these signals, the AP feeds the perturbed CSI signatures to its DL localization model, which will consequently output a wrong estimation that is desired by the attacker. The main advantages of FooLoc are that it has small perturbations with respect to original signals and remains unharmful to message demodulation at the AP.

As shown in Fig.~\ref{fig:systemoverview}, the core components of FooLoc include \textit{Over-The-Air Perturbation Design} and \textit{Adversarial Weight Optimization}.
\begin{itemize}
	\item \textbf{Over-The-Air Perturbation Design.} First, we investigate the multiplicative and repetitive properties of over-the-air perturbations and formalize their impacts on uplink CSI measurements. Then, we define the notions of adversarial examples as well as targeted and untargeted adversarial attacks on wireless localization. Additionally, we prove that our adversarial perturbation remains unharmful to the payload decoding at the AP.
	\item \textbf{Adversarial Weight Optimization.} First, we detail our attack strategy and propose a generalized objective function that integrates both targeted and untargeted attacks. Then, adversarial attacks on DL localization models are formulated as a box-constrained problem that minimizes the objective function while satisfying the constraints of robustness, imperceptibility as well as efficiency. Moreover, we carefully transform the above constrained problem into an equivalent unconstrained one for easing the difficulty of problem optimization.
\end{itemize}

\subsection{Over-The-Air Perturbation Design}
In this subsection, we first investigate the unique multiplicative and repetitive properties of over-the-air perturbations and define adversarial examples in indoor localization.

\textbf{Multiplicative Property.} Most of the prior studies on wireless adversarial attacks synthesize an adversarial example $ x^{ad} $ for each genuine sample $ x $ using an additive perturbation $ r $ likewise generating adversarial images in the computer vision domain as $ x^{ad} = x + r $. However, it is inapplicable for performing over-the-air attacks in real-world wireless channels. In over-the-air attacks, the attacker can change model inputs only via multiplicative perturbations. The reason stems from the fact that a received signal is the product of a channel response and a transmitted signal in the frequency domain~\cite{tse2005fundamentals}. Hence, one uplink CSI measurement has a proportional relationship with the perturbed training signals as indicated in Eq.~\eqref{eq: csi estimation}.

\begin{figure}
	\centering
	\includegraphics[width=1\linewidth]{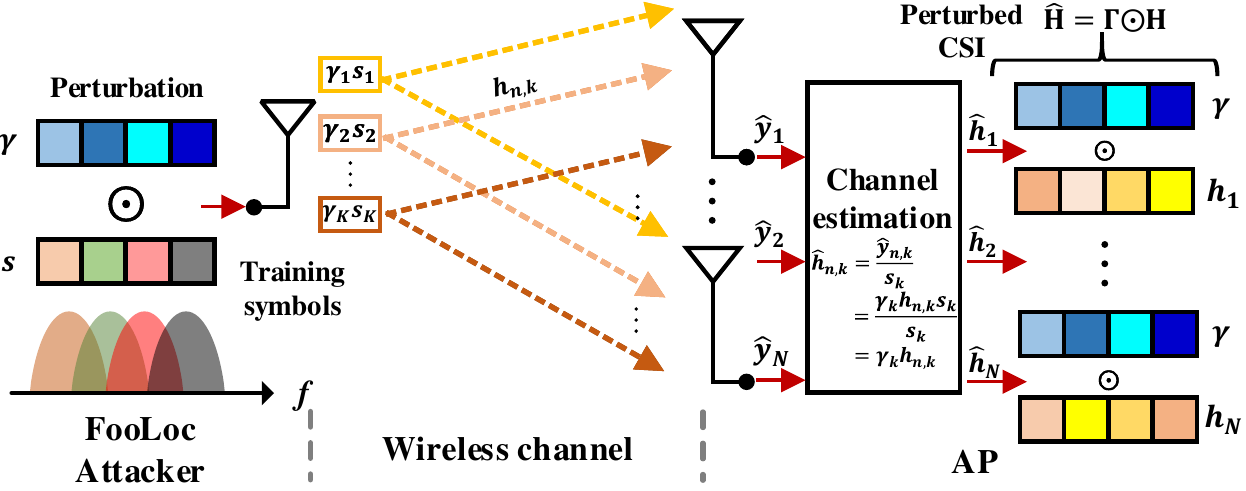}
	\caption{Illustration of over-the-air adversarial perturbations from the attacker to the victim AP.}
	\label{fig:otaattack}
\end{figure}

Specifically, as depicted in Fig.~\ref{fig:otaattack}, when attempting to launch over-the-air attacks, FooLoc first generates a real-valued multiplicative perturbation set $ \boldsymbol{\gamma}   = [\gamma_1, \cdots, \gamma_k, \cdots, \gamma_K] \in \mathbb{R}^{1 \times K} $ for its $ K $-element training sequence $ \mathbf{s} = [s_1, \cdots, s_k, \cdots, s_K] \in \mathbb{C}^{1 \times K}  $, which is known by the AP. Then, the scaled sequence $ \mathbf{s}_t \in \mathbb{C}^{1 \times K} $ can be obtained as 
\begin{align}
	\mathbf{s}_t = \boldsymbol{\gamma} \odot \mathbf{s} = [\gamma_1 s_1, \cdots, \gamma_k s_k, \cdots, \gamma_K s_K],
\end{align}
where $ \odot $ is the Hadamard product for element-wise production. Then, FooLoc transmits $ \mathbf{s}_t $ to the victim AP over realistic wireless channels. When hearing the signal, the AP with $ N $ antennas receives a measurement $ \hat{\mathbf{Y}} \in \mathbb{C}^{N \times K} $ and estimates their uplink channel $ \hat{\mathbf{H}} \in \mathbb{C}^{N \times K} $ using Eq.~\eqref{eq: csi estimation}. Therein, each entry of $ \hat{\mathbf{H}} $ can be denoted as $ \hat{h}_{n,k} $, representing the perturbed channel response between the client and the $ n $-th AP antenna at the $ k $-th subcarrier. Let us assume that the corresponding true channel estimation is $ \mathbf{H} \in \mathbb{C}^{N \times K} $ with each entry denoted as $ h_{n,k} $. From Eq.~\eqref{eq: csi estimation}, we can have
\begin{align} \label{eq: perturbed channel}
\hat{h}_{n,k} = \frac{\hat{y}_{n,k}}{s_k}=\frac{ h_{n,k} \gamma_k s_k}{s_k} = \gamma_k h_{n,k}.
\end{align}
According to Eq.~\eqref{eq: perturbed channel}, we can see that $ h_{n,k} $, as the original channel response, is proportionally perturbed by $ \gamma_k $, suggesting that over-the-air perturbations have a multiplicative effect on uplink CSI measurements.

Using such multiplicative weights, FooLoc can easily manipulate uplink CSI measurements through the standard channel estimation process as depicted in Fig.~\ref{fig:otaattack}, which lays the foundation for further over-the-air attacks.

\textbf{Repetitive Property.} Given the multiplicative perturbation, we proceed to investigate the unique pattern of our perturbation weights received by the AP. Existing studies on adversarial attacks create different perturbation weights for different input elements. Yet, this is not the case for adversarial attacks over wireless channels. 

As illustrated in Fig.~\ref{fig:otaattack}, the uplink transmission from the attacker to the AP can be modeled as a single-input-multiple-output (SIMO) channel, which suggests a one-to-many relationship between the elements of one perturbation $ \boldsymbol{\gamma} $ and the perturbed CSI measurement $ \hat{\mathbf{H}} $. Mathematically, given the perturbation weight $ \gamma_k $, the $ k $-th column of $ \hat{\mathbf{H}} $ represents all estimated channel responses for the $ k $-th subcarrier and can be further written as
\begin{align} 
	\hat{\mathbf{h}}_{k} =  \begin{bmatrix} \hat{h}_{1,k} \\ \vdots \\ \hat{h}_{N,k} \end{bmatrix} =    \begin{bmatrix} \gamma_k h_{1,k} \\ \vdots \\ \gamma_k h_{N,k} \end{bmatrix} =   \gamma_k \begin{bmatrix}  h_{1,k} \\ \vdots \\ h_{N,k} \end{bmatrix}.
\end{align} 
The above equation shows that all receiving antennas share the same perturbation weight with respect to each subcarrier. Hence, the overall received perturbation weights $ \mathbf{\Gamma} \in \mathbb{R}^{N \times K} $ on $ \hat{\mathbf{H}} $ have a repetitive pattern as
\begin{align} \label{eq: pertubation weights}
	\boldsymbol{\Gamma}  = \mathit{J}_{\scriptscriptstyle N \times 1} \otimes \boldsymbol{\gamma}  = \begin{bmatrix} \gamma_{1} & \cdots & \gamma_{K}  \\ \vdots& \vdots& \vdots\\  \gamma_{1}&\cdots& \gamma_{K}  \end{bmatrix},
\end{align}  
where $ \mathit{J}_{\scriptscriptstyle N \times 1} $ is the all-ones matrix with a size of $ N \times 1 $ and $ \otimes $ denotes the Kronecker product that helps $ \boldsymbol{\gamma} $ expanding in the vertical dimension in Eq.~\eqref{eq: pertubation weights}.

With the observations of multiplicative weights and repetitive patterns, we can finally formulate the impact of FooLoc's perturbations on uplink CSIs as 
\begin{align} \label{eq: attacking formulation}
	\hat{\mathbf{H}}  = \mathit{J}_{\scriptscriptstyle N \times 1} \otimes \boldsymbol{\gamma} \odot \mathbf{H}.
\end{align}

\textbf{Adversarial Perturbations.} Next, we define the notion of over-the-air adversarial examples in the context of indoor localization. Let $ \mathcal{P} \in \mathbb{R}^{2} $ be the 2D area, where the AP provides wireless connectivity. We denote $ f_{\theta}(\cdot ): \mathcal{X} \rightarrow  \mathcal{P} $ as the localization DNN used by the AP, where $ \theta $ stands for the already trained parameters using uplink CSI fingerprints $ \mathcal{X}^{u}_{\mathcal{A}}$ that are collected at a set of reference spots $ \mathcal{A}  \subset \mathcal{P} $. Therein, each input sample $ \mathbf{X}^{u} \in \mathbb{R}^{N \times K} $ represents the amplitudes of one uplink CSI. Moreover, we assume that our attacker locates at a location $ \mathbf{p} \in  \mathcal{P} $, i.e., the genuine spot, and manipulates its uplink channel using a perturbation $ \boldsymbol{\gamma}_p $. Considering that amplitude features are essentially the absolute values of complex-valued channel responses, the real-valued perturbation weights in Eq.~\eqref{eq: attacking formulation} will have the same linear scaling effect on corresponding CSI amplitudes. Using this property, we can derive our adversarial example $ \hat{\mathbf{X}}^{u}_p $ as 
\begin{align} \label{eq: adversarial example}
	\hat{\mathbf{X}}^{u}_p = \mathit{J}_{\scriptscriptstyle N \times 1} \otimes \boldsymbol{\gamma}_p \odot \mathbf{X}^{u}_p,
\end{align}
where $ \mathbf{X}^{u}_p $ represents the true uplink CSI amplitudes.

Based on the above notion of adversarial examples, we further define the adversarial perturbations for targeted and untargeted attacks, respectively, on indoor localization DNNs. In the targeted case, one successful perturbation $ \boldsymbol{\gamma}_p $ would mislead a location estimate $ f_{\theta}( \hat{\mathbf{X}}^{u}_p ) $ to a targeted spot $ \mathbf{q} \in  \mathcal{P} $ as close as possible, where $ \mathbf{q} \neq \mathbf{p} $. That is, we seek a perturbation $ \boldsymbol{\gamma}_p $ such that
\begin{align} \label{eq: targeted attack}
	\mathcal{D}\left( f_{\theta} \left(  \mathit{J}_{\scriptscriptstyle N \times 1} \otimes \boldsymbol{\gamma}_p \odot \mathbf{X}^{u}_p \right) , \mathbf{q}\right)   \le d_{max} ,
\end{align}
where $ \mathcal{D} (\cdot,\cdot) $ is the euclidean distance and $ d_{max} $ represents the acceptable maximal distance error. Whereas, in the untargeted case, one adversarial perturbation $ \boldsymbol{\gamma}_p $ would make $ f_{\theta}( \hat{\mathbf{X}}^{u}_p ) $ away from the genuine location $ \mathbf{p} $ as far as possible. Similarly, given the acceptable minimal distance error $ d_{min} $, we expect a perturbation $ \boldsymbol{\gamma}_p $ satisfying
\begin{align} \label{eq: untargeted attack}
\mathcal{D}\left( f_{\theta} \left(  \mathit{J}_{\scriptscriptstyle N \times 1} \otimes \boldsymbol{\gamma}_p \odot \mathbf{X}^{u}_p \right) , \mathbf{p}\right)   \ge d_{min}.
\end{align}
We will specify the configurations of two acceptable distance errors $ d_{min} $ and $ d_{max} $ and verify the validity of such configurations in our experiments.

\textbf{Impact on Message Demodulation.} One of the major benefits of our multiplicative perturbation $ \boldsymbol{\gamma} $ defined in Eq.~\eqref{eq: attacking formulation} is that it has no impact on message demodulation at the AP. Specifically, in each packet transmission, FooLoc not only applies the multiplicative perturbations on pre-defined LTF symbols $ \mathbf{s} $, but also uses them accordingly on the subsequent payload signal $ \mathbf{u} = [u_1, \cdots, u_k, \cdots, u_K] \in \mathbb{C}^{1 \times K} $. After that, the perturbed payload will go through the same real channel as the perturbed training sequence. In this way, although the AP obtains a fake CSI response, the original message is perturbed in the same way. Thus, based on the perturbed response $ \hat{h}_{n,k} $ in Eq.~\eqref{eq: perturbed channel}, the payload signal $ u_k $ still can be correctly decoded from the received signals $ h_{n,k} \gamma_k u_k $. This process can be mathematically expressed as
\begin{align}
	\frac{  h_{n,k} \gamma_k u_k}{\hat{h}_{n,k}} = \frac{h_{n,k} \gamma_k u_k}{\gamma_k h_{n,k}} = u_k.
\end{align}
Hence, our adversarial perturbations remain unharmful to the message transmission from the attacker to the AP. The only impact of such perturbations is that the AP feeds falsified CSIs to its localization DNN.

\subsection{Adversarial Weight Optimization}
In this subsection, we first detail our attack strategy and formulate adversarial perturbation generation as a box-constrained optimization problem. Then, we transform it into an unconstrained one.

\textbf{Attack Strategy.} Since uplink CSI measurements are unknown to the attacker, one possible attack strategy is to blindly manipulate its LTF symbols in a brute-force manner. However, such an approach is prohibitively inefficient and time-consuming. Instead of blindly searching, FooLoc exploits the accessible and informative downlink CSI measurements, which can be easily obtained from the AP's beacon or ACK packets in Wi-Fi networks~\cite{xiao2005ieee}. Concretely, when our attacker stays at the genuine spot $ \mathbf{p} $, it first collects some downlink CSI measurements and obtains a set of amplitude features $ \mathcal{X}^{d}_{p} $, where $ \mathbf{X}^{d}_p \in \mathbb{R}^{N \times K} $. Then, FooLoc simulates the over-the-air attacks using Eq.~\eqref{eq: adversarial example} and optimizes the perturbation weights based on $ \mathcal{X}^{d}_{p} $. After that, it multiplies the optimized weights $ \boldsymbol{\gamma}_p $ with the pre-defined training sequence $ \mathbf{s} $ and sends their product results to the AP for attacking its localization model $ f_{\theta}(\cdot ) $. Because uplink and downlink channel responses are similar as aforementioned, the perturbation weights learned from downlink CSI measurements are expected to generalize well to uplink ones.

\textbf{Problem Formulation.} With the above attack strategy, we first integrate both targeted attacks~\eqref{eq: targeted attack} and untargeted attacks~\eqref{eq: untargeted attack} in wireless localization into one objective function $\mathcal{J} \left( \boldsymbol{\gamma}_p, f_{\theta}  \right) $ as 
\begin{align} \label{eq: targeted objective function}
\notag \mathcal{J} \left( \boldsymbol{\gamma}_p, f_{\theta}  \right) \triangleq & (1-\omega)   \mathbb{E}_{\mathcal{X}^{d}_{p}}  \left[  \mathcal{D} \left(f_{\theta} \left(  \mathit{J}_{\scriptscriptstyle N \times 1} \otimes \boldsymbol{\gamma}_p \odot \mathbf{X}^{d}_p \right), \mathbf{q} \right) - d_{max} \right]^{+} \\& + \omega  \mathbb{E}_{\mathcal{X}^{d}_{p}}  \left[ d_{min} - \mathcal{D} \left( f_{\theta} \left(  \mathit{J}_{\scriptscriptstyle N \times 1} \otimes \boldsymbol{\gamma}_p \odot \mathbf{X}^{d}_p \right), \mathbf{p} \right) \right]^{+}.
\end{align}
Therein, $ \omega $ indicates the attack type and takes values in the set $ \left\lbrace 0,1 \right\rbrace  $, where $ \omega=0 $ stands for targeted attacks and $ \omega=1 $ is for untargeted attacks. $ \mathbb{E}_{\mathcal{X}^{d}_{p}} [\cdot] $ is the expectation over the dataset $ \mathcal{X}^{d}_{p} $ and $ [a]^{+} =  \max(a,0)  $ denotes the positive part of $ a $. 

Using this objective function, we formulate the problem of adversarial attacks on the localization model $ f_{\theta}(\cdot ) $ as
\begin{align}
\underset{\boldsymbol{\gamma}_p }{\text{minimize}} \ &   \mathcal{J} \left( \boldsymbol{\gamma}_p, f_{\theta} \right) + \beta \; \norm{\Delta \boldsymbol{\gamma}_p }_{2},  \label{eq: white-box attack objective} \\
 \text{subject to} \ &  \norm{\boldsymbol{\gamma}_p -\mathit{J}_{\scriptscriptstyle 1 \times K}}_{\infty} < \delta_{max} < 1. \label{eq: white-box attack constraint}
\end{align}
Therein, $ \Delta\boldsymbol{\gamma_p} = \left[  \gamma_{p,i} - \gamma_{p,i-1}\right]_{i = 2,\cdots,K}  $ is the difference vector of $ \boldsymbol{\gamma}_p $ and $ \beta $ denotes a hyperparameter. In addition, $ \norm{\mathbf{a}}_{2} $ is the $ \mathit{l}_{2} $ norm and $ \norm{\mathbf{a}}_{\infty} = \max\left(  |a_1|, \cdots, |a_n| \right)     $ is the $ \mathit{l}_{\infty} $ norm. In the following, we explain the design rationale of the above box-constrained problem.

\textit{Robustness.} When $ \omega=0 $ in the objective function $\mathcal{J} \left( \cdot  \right) $, we minimize the average error between the distance $ \mathcal{D} \left(f_{\theta}( \hat{\mathbf{X}}^{d}_p), \mathbf{q} \right) $ and the threshold $ d_{max} $ over the entire downlink CSI dataset $ \mathcal{X}^{d}_{p} $. This is because due to the random nature of environmental noise in Wi-Fi CSI signatures, two CSI instances from one spot are unlikely to be exactly the same. As a consequence, the perturbation that is crafted for a specific CSI sample may have little effect on another one with a high probability. To boost the robustness of our adversarial perturbations, FooLoc seeks a universal perturbation that causes all the samples in $ \mathcal{X}^{d}_{p} $ to be estimated at a neighboring area of the targeted location $ \mathbf{q} $. The same reason holds for the untargeted attacks when $ \omega=1 $.

\begin{figure}
	\centering
	\includegraphics[width=1\linewidth]{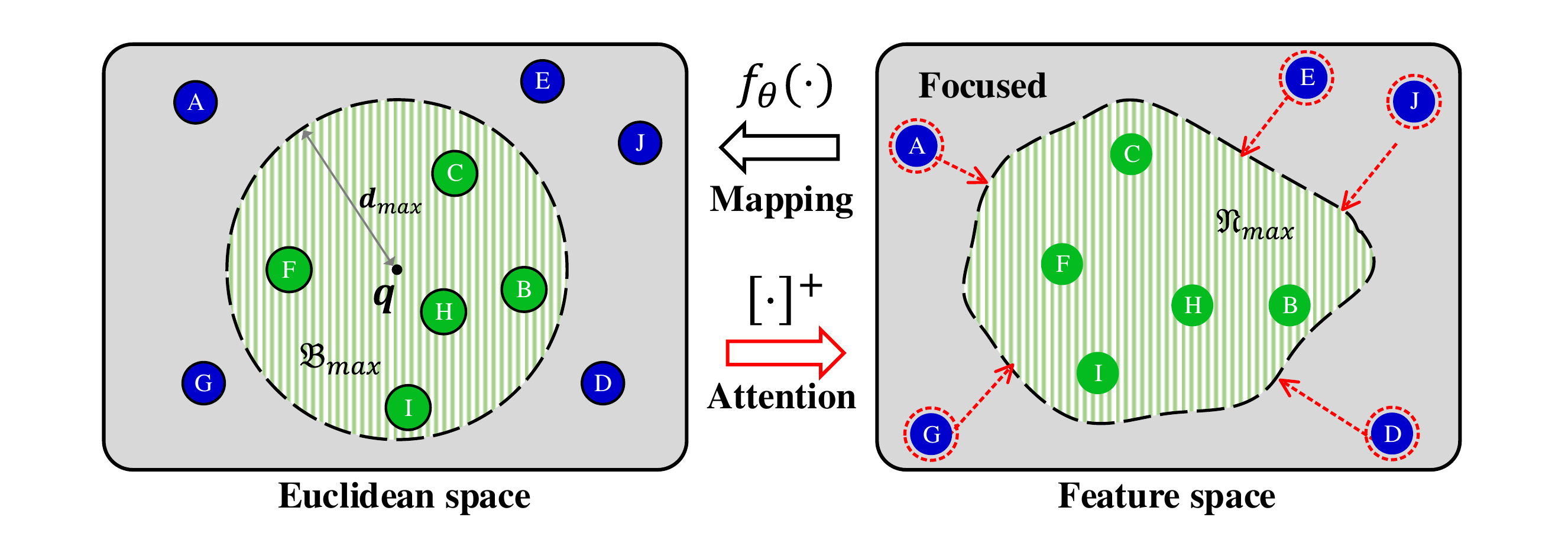}
	\caption{Illustration of FooLoc's attention scheme for targeted attacks during perturbation optimization.}
	\label{fig:attention}
\end{figure}

\textit{Imperceptibility.} The second term in Eq.~\eqref{eq: white-box attack objective} and the constraint in Eq.~\eqref{eq: white-box attack constraint} together guarantee the imperceptibility of our adversarial perturbations. Specifically, $ \norm{\Delta \boldsymbol{\gamma}_p }_{2} $ quantifies the smoothness of one perturbation $ \boldsymbol{\gamma}_p $ by measuring the difference between its consecutive weights. The smaller the difference, the smoother the perturbation. In the extreme case $ \norm{\Delta \boldsymbol{\gamma}_p }_{2} = 0 $, $ \boldsymbol{\gamma}_p $ shall be a constant. In this condition, $ \boldsymbol{\gamma}_p $ has the same linear scaling effect on each element of one CSI measurement and can not manipulate its changing trends. Moreover, the constraint~\eqref{eq: white-box attack constraint} limits the perturbation strength and makes sure that FooLoc always searches a perturbation $ \boldsymbol{\gamma}_p $ within the $ \mathit{l}_{\infty} $ norm ball with a radius $ \delta_{max} $ centering at $ \mathit{J}_{\scriptscriptstyle 1 \times K} $ during optimization process. The choose of $ \mathit{l}_{\infty} $ norm in Eq.~\eqref{eq: white-box attack constraint} makes each adversarial weight $ \gamma_{p,k} $ in $ \boldsymbol{\gamma}_p $ satisfying $ 1-\delta_{max} < \gamma_{p,k} < 1 + \delta_{max} $. The above two designs can guarantee a minimally-perturbed signal $\hat{\mathbf{X}}^{d}_p $ that is seemingly alike to the original signal $ \mathbf{X}^{d}_p$ when received by the AP. 

\textit{Efficiency.} At each optimization step, not all samples are necessary for updating perturbation weights. Without loss of generality, we take $ \omega=0 $, i.e., the targeted attacks, for explaining. Let $ \mathfrak{R}_{max}  \triangleq \left\lbrace \mathbf{X}: \mathcal{D} \left(  f_{\theta} \left( \mathbf{X} \right), \mathbf{q} \right)  < d_{max}  \right\rbrace   $ be the set of amplitude features, whose location estimates are within $ \mathfrak{B}_{d_{max}}(\mathbf{q}) \subset  \mathcal{P} $, i.e., the ball with a radius of $ d_{max} $ centering at the targeted spot $  \mathbf{q} $ in the Euclidean space. After some optimization steps, a part of perturbed CSI samples may have already been mapped in $ \mathfrak{B}_{d_{max}}(\mathbf{q})  $ by $ f_{\theta}(\cdot) $, i.e., the green circles in the Euclidean space in Fig.~\ref{fig:attention}. In this condition, these samples are unnecessary for optimizing new perturbation weights in the next step. Based on this observation, we devise an attention scheme to enhance the efficiency of our optimization problem. In particular, FooLoc uses the operator $ [\cdot]^{+} $ in $\mathcal{J} \left( \boldsymbol{\gamma}_p, f_{\theta}  \right) $ to discriminate whether location estimates are inside or outside of $ \mathfrak{B}_{d_{max}}(\mathbf{q})  $. Then, it strategically pays attention to outside samples and ignores inside ones. This operation will generally decrease the number of needed samples at each optimization step and thus lead to a lower overall computational overhead. 

\begin{figure}[t]
	\centering
	\includegraphics[width=0.9\linewidth]{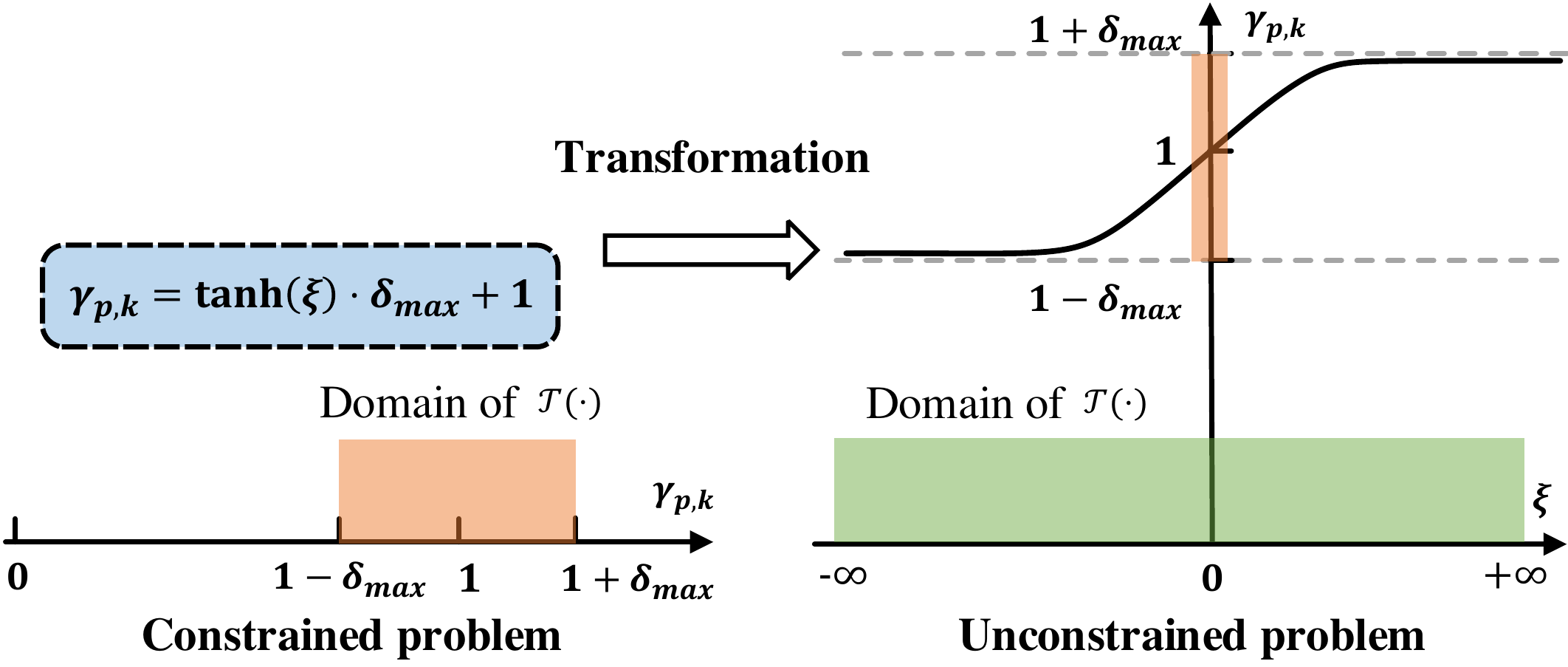}
	\caption{Illustration of weight transformation in problem optimization. For simplicity, we take one element of $ \boldsymbol{\gamma}_p $ for illustration.}
	\label{fig:transformation}
\end{figure}

\textbf{Problem Optimization.} With the optimization problem~\eqref{eq: white-box attack objective}, we proceed to design a dedicated optimization scheme for generating our adversarial perturbations. Because our perturbations are multiplicative rather than additive, traditional perturbation generation algorithms, such as the well-known fast gradient sign method (FGSM)~\cite{goodfellow2014explaining}, are inapplicable for our optimization problem. Thus, we need to directly solve the problem~\eqref{eq: white-box attack objective} using other general gradient based optimization methods, such as stochastic gradient descent (SGD) and adaptive moment estimation (Adam). However, the constraint term~\eqref{eq: white-box attack constraint} restricts the domain of the objective function $ \mathcal{J} \left( \cdot \right) $ in the space $ \left(  1 - \delta_{max}, 1 + \delta_{max} \right) ^{1 \times K}  $ and makes the optimization problem as a box-constrained one, which is not naively supported by such gradient-based optimization methods.

\begin{algorithm}[t] 
	\caption{Over-the-air adversarial attacks on DL localization models. } 
	\label{adversarial algorithm} 
	\begin{algorithmic}
		\STATE \textbf{Input:} Downlink CSI samples $ \mathcal{X}^{d}_{p}  $, the DL localization model  $ f_{\theta}(\cdot )  $, the genuine and targeted spots $ \left\lbrace  \mathbf{p}, \mathbf{q} \right\rbrace$, the acceptable distance errors $\left\lbrace d_{min}, d_{max}\right\rbrace  $ and the attack type $ \omega $
		\STATE \begin{center}
			$ \boldsymbol{\xi} \gets \text{random}(1, K) \in \mathbb{R}^{1 \times K} $ $ \blacktriangleright $ initialization \\
		\end{center}
		\FOR{the number of training iterations}
		\STATE Sample a mini-batch of training data $ \left\lbrace  \mathbf{X}^{d}_{p,i}  \right\rbrace^M_{i=1}  $ from $ \mathcal{X}^{d}_{p}  $
		\STATE Generate adversarial examples
		\STATE \begin{center}
			$ \boldsymbol{\gamma}_p \gets \tanh\left( \boldsymbol{\xi} \right) \cdot \delta_{max} + \mathit{J}_{\scriptscriptstyle 1 \times K} $\\
			$ \hat{\mathbf{X}}^{d}_{p,i} \gets \mathit{J}_{\scriptscriptstyle N \times 1} \otimes \boldsymbol{\gamma}_p  \odot \mathbf{X}^{d}_{p,i} $
		\end{center}
		\STATE Update parameters $ \boldsymbol{\xi} $:
		\IF{$ \omega = 0 $} 
		\STATE \begin{center}
			$ \boldsymbol{\xi} \gets \boldsymbol{\xi} - \eta \nabla_{\boldsymbol{\xi}}   \left[ \frac{\sum_{i}^{M}\left[  \mathcal{D} \left(f_{\theta} \left(  \hat{\mathbf{X}}^{d}_{p,i} \right), \mathbf{q} \right) - d_{max} \right]^{+}}{M}  + \beta \; \norm{\Delta \boldsymbol{\gamma}_p }_{2} \right] $
		\end{center}
		\ENDIF
		\IF{$ \omega = 1 $} 
		\STATE \begin{center}
			$ \boldsymbol{\xi} \gets \boldsymbol{\xi} - \eta \nabla_{\boldsymbol{\xi}} \left[  \frac{\sum_{i}^{M} \left[ d_{min} - \mathcal{D} \left(f_{\theta} \left(  \hat{\mathbf{X}}^{d}_{p,i} \right), \mathbf{p} \right)  \right]^{+}}{M}  + \beta \; \norm{\Delta \boldsymbol{\gamma}_p }_{2} \right] $
		\end{center}
		\ENDIF
		\ENDFOR
		\STATE Generate and transmit perturbed LTFs and payload signals
		\begin{center}
			$  \mathbf{s}_t \gets \left( \tanh\left( \boldsymbol{\xi} \right) \cdot \delta_{max} + \mathit{J}_{\scriptscriptstyle 1 \times K} \right)  \odot \mathbf{s} $ \\	$  \mathbf{u}_t \gets \left(\tanh\left( \boldsymbol{\xi} \right) \cdot \delta_{max} + \mathit{J}_{\scriptscriptstyle 1 \times K} \right)  \odot \mathbf{u} $
		\end{center}
	\end{algorithmic} 
\end{algorithm}

To deal with this issue, we transform the box-constrained problem~\eqref{eq: white-box attack objective} into an equivalent unconstrained one for easing its optimization difficulty. To do this, we first make $ \boldsymbol{\gamma}_p $ satisfying the constraint~\eqref{eq: white-box attack constraint} via the transformation as
\begin{align}\label{eq: transformation}
	\boldsymbol{\gamma}_p = \tanh\left( \boldsymbol{\xi} \right) \cdot \delta_{max} + \mathit{J}_{\scriptscriptstyle 1 \times K},
\end{align}
where $ \boldsymbol{\xi} \in \mathbb{R}^{1 \times K} $. Moreover, $ \tanh\left( x \right) = \frac{e^{x}-e^{-x}}{e^{x}+e^{-x}} $ is the hyperbolic tangent function with the range $ \left(-1, 1 \right)  $. As illustrated in Fig.~\ref{fig:transformation}, each element $ \gamma_{p,k} $ in $ \boldsymbol{\gamma}_p $ is naturally confined to the interval $ \left(   1-\delta_{max}, 1 + \delta_{max} \right)   $ using the above transformation, which is equivalent to the constraint $ \norm{\boldsymbol{\gamma}_p -\mathit{J}_{\scriptscriptstyle 1 \times K}}_{\infty} < \delta_{max} $. Then, we substitute $ \boldsymbol{\gamma}_p $ with Eq.~\eqref{eq: transformation} in the original problem~\eqref{eq: white-box attack objective}, which will convert the domain of $ \mathcal{J} \left( \cdot \right) $ into the space $ \mathbb{R}^{1 \times K}  $. In this way, we obtain an equivalent unconstrained problem of adversarial perturbation generation as
\begin{align}\label{eq: equivalent problem}
	\underset{ \boldsymbol{\xi} \in \mathbb{R}^{1 \times K}  }{\text{minimize}} \; & \mathcal{J} \left( \boldsymbol{\gamma}_p, f_{\theta} \right) + \beta \norm{\Delta \boldsymbol{\gamma}_p }_{2},\\
	\text{where} & \; \boldsymbol{\gamma}_p = \tanh\left( \boldsymbol{\xi} \right) \cdot \delta_{max} + \mathit{J}_{\scriptscriptstyle 1 \times K}.
\end{align}
In this condition, we can leverage traditional gradient-based methods to solve the optimization problem~\eqref{eq: equivalent problem}.

At last, FooLoc can apply the well-trained adversarial weights on pre-defined LTF symbols as well as payload signals and transmit their product results over wireless channels to fool the localization DNN $ f_{\theta}(\cdot ) $ at the AP. The way to launch our over-the-air adversarial attacks is summarized in Algorithm~1. In our experiments, we empirically set $ \delta_{max} = 0.15 $ and use the SGD optimizer for searching optimal perturbation weights. 

\section{Evaluation} \label{sec:experiment}

\subsection{Victim DNNs and Evaluation Metrics}

\textbf{Victim DNNs.} To evaluate FooLoc, we build two victim localization models, i.e., $ \textit{DNN}_{A} $ and $ \textit{DNN}_{B} $, using mainstream neural network architectures. In particular, both $ \textit{DNN}_{A} $ and $ \textit{DNN}_{B} $ are set as regression models, which take raw multi-dimensional CSI samples as inputs and output a continuous-valued location estimate. The structures and parameters of two DNNs are present in Table~\ref{tab: dnn structure}. As the table shows, $ \textit{DNN}_{A} $ is a fully connected neural network (FCNN). It first normalizes each sample element into the interval $ [0, 1] $ along the antenna dimension for effective inference~\cite{goodfellow2016deep} and flattens a normalized sample into a one-dimensional tensor. Then, $ \textit{DNN}_{A} $ leverages six fully connected (fc) layers to extract hidden features and predicts the corresponding device location. $ \textit{DNN}_{B} $ is a convolutional neural network (CNN) and consists of three convolutional (conv) layers and three fully connected layers. It also performs data normalization before feeding CSI samples into its convolutional layers. In addition, we build $ \textit{DNN}_{A} $ and $ \textit{DNN}_{B} $ on the PyTorch framework.
\begin{table}
	\centering
	\caption{The structures and Parameters of Victim DNNs Used in Our Experiments.}\label{tab: dnn structure}
	\begin{tabular}{|c|c|c|c|}
		\hline
		\multicolumn{2}{|c|}{}               & $ \textit{DNN}_{A} $                  & $ \textit{DNN}_{B} $         \\ \hline
		\multicolumn{2}{|c|}{Pre-processing} & Normalize\&Flatten & Normalize    \\ \hline
		\multirow{7}{*}{Layers}     & \#1    & fc1024, Linear            & conv256@1$\times$1, ReLu \\ \cline{2-4} 
		& \#2    & fc512, ReLu          & conv128@1$\times$1, ReLu \\ \cline{2-4} 
		& \#3    & fc1024, Linear          & conv128@1$\times$1, ReLu \\ \cline{2-4} 
		& \#4    & fc512, ReLu          & fc512, ReLu \\ \cline{2-4} 
		& \#5    & fc1024, Linear          & fc256, ReLu \\ \cline{2-4} 
		& \#6    & fc2, Sigmoid          & fc2, Sigmoid    \\ \cline{2-4}  \hline
	\end{tabular}
\end{table}

\textbf{Evaluation Metrics.} We use the following metrics to measure FooLoc's performance.
\begin{itemize}
	\item \textbf{Localization Error (LE).} Given a localization model  $ f_{\theta}(\cdot ) $ and an input sample $ \mathbf{X}^{u}_{g} $ from the ground-truth spot $ \mathbf{g} $, the LE to $ \mathbf{g} $ is computed as 
	\begin{align}
	\notag \mathcal{D} \left( f_{\theta}(\mathbf{X}^{u}_{g} ), \mathbf{g} \right) =	\norm{f_{\theta}(\mathbf{X}^{u}_{g} )-\mathbf{g}}_2.
	\end{align}
	\item \textbf{Attack Success Rate (ASR).} Given a set of perturbed uplink CSIs $ \hat{\mathbf{X}}^{u}_p = \left\lbrace \hat{\mathbf{X}}^{u}_{p,m} \right\rbrace_{m=1:M} $ pertaining to the attacker's true spot $ \mathbf{p} $ and an adversarial perturbation $ \boldsymbol{\gamma}_p $, the ASR of targeted attacks with a targeted spot $ \mathbf{q} $ is
	\begin{align}
	\notag	\sum_{m} \mathbbm{1} \left( \mathcal{D} \left(f_{\theta} \left( \hat{\mathbf{X}}^{u}_{p,m} \right), \mathbf{q} \right) - d_{max} \leq 0 \right) /M, 
	\end{align}
	where $ \mathbbm{1}(\cdot) $ denotes the indication function and $ d_{max} $ is the acceptable maximal distance error. It represents the probability that a perturbed location estimation $ f_{\theta} \left( \hat{\mathbf{X}}^{u}_{p,m} \right) $ is inside the ball centering at the targeted spot $ \mathbf{q} $ with a radius of $  d_{max} $. Similarly, the ASR of untargeted attacks is given as
	\begin{align}
	\notag	\sum_{m} \mathbbm{1} \left( \mathcal{D} \left(f_{\theta} \left(  \hat{\mathbf{X}}^{u}_{p,m} \right), \mathbf{p} \right) - d_{min} \geq 0 \right) /M,
	\end{align}
	where $ d_{min} $ is the acceptable minimal distance error. It indicates the probability that a perturbed location estimation $ f_{\theta} \left( \hat{\mathbf{X}}^{u}_{p,m} \right) $ is at the outside of the ball centering at the true spot $ \mathbf{p} $ with a radius of $ d_{min} $.
	\item \textbf{Perturbation-To-Signal Ratio (PSR).} Given the perturbed uplink CSI $ \hat{\mathbf{X}}^{u}_p $ and corresponding original one $ \mathbf{X}^{u}_{p} $ at the genuine spot $ \mathbf{p} $, the PSR is computed as
	\begin{align}
	\notag	\text{PSR}=20 \log_{10} \frac{\norm{\hat{\mathbf{X}}^{u}_p-\mathbf{X}^{u}_p}_2}{\norm{\mathbf{X}^{u}_p}_2}.
	\end{align}
\end{itemize}

\begin{figure}
	\centering
	\includegraphics[width=0.9\linewidth]{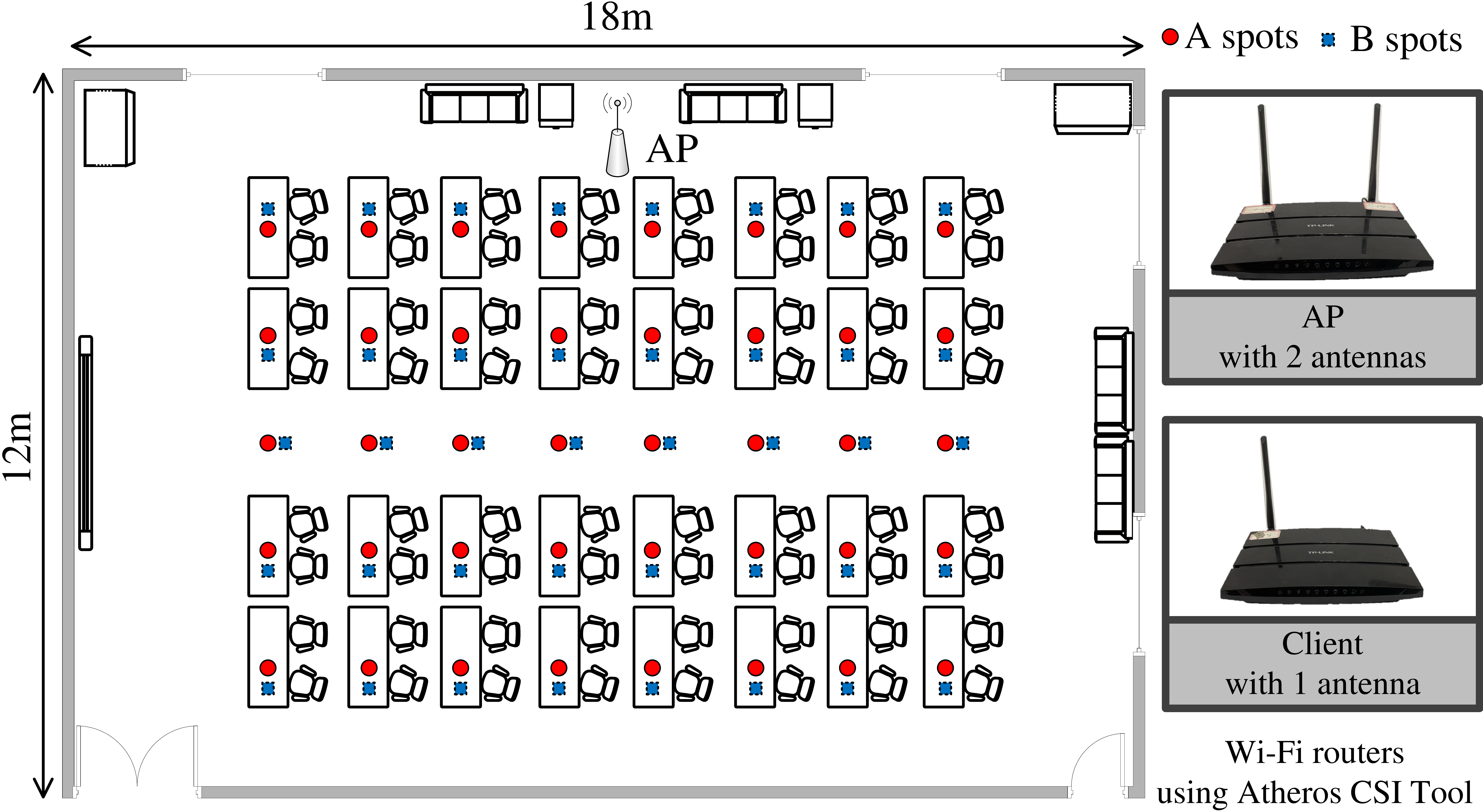}
	\caption{Floor plan of the experiment environment and experimental platform in offline experiments.}
	\label{fig:floorplan}
\end{figure}

\subsection{Offline Experiments}
In this subsection, we conduct our offline experiments, in which both uplink and downlink CSI measurements are first collected in real-world environments. In this setting, the attacker optimizes adversarial perturbations using downlink CSIs and then applies the learned perturbations directly on the collected uplink ones based on Eq.~\eqref{eq: adversarial example} to spoof localization DNNs. 

\textbf{Implementation.} In offline experiments, we implement FooLoc using two TL-WDR4310 Wi-Fi routers and one Lenovo laptop. Specifically, one router with two antennas is fixed at one spot to act as an AP, and the left one is equipped with one antenna to work as a mobile client to communicate with the AP from different spots. Moreover, we connect the laptop with two routers via Ethernet cables and run Atheros CSI Tool~\cite{xie2015precise}. Using this tool, each router is set to work at the 2.4~GHz Wi-Fi band and record channel responses of 56 subcarriers. Hence, one CSI sample has a size of $ 1\times 2 \times 56 $. 

\textbf{Data Collection.} We collect CSI measurements in a $ 12\times 18 \, \text{m}^{2} $ meeting room as shown in Fig.~\ref{fig:floorplan}. The AP is placed at one end of the room to avoid isotropy for better localization performance~\cite{wang2016csi,wang2015deepfi}. We move the client among 40 selected locations with a spacing distance of 1.5~m, i.e., A spots in Fig.~\ref{fig:floorplan}, to collect uplink CSI measurements at the AP. Accordingly, we choose 40 locations around A spots, i.e., B spots in Fig.~\ref{fig:floorplan}, to record uplink and downlink CSI measurements, respectively. At each spot, 250 CSI samples are recorded during data collection. Thus, we can obtain three datasets $ \mathcal{D}_{A} $, $ \mathcal{D}_{B} $ and $ \mathcal{D}_{C} $. In particular, $ \mathcal{D}_{A} $ includes 10K uplink CSI samples from A spots and is used for training localization DNNs at the AP. $ \mathcal{D}_{B} $ consists of 10K downlink samples from B spots and is used by the attacker to generate adversarial perturbations. $ \mathcal{D}_{C} $ has 10K uplink samples from B spots and is responsible for testing FooLoc.

\begin{figure}
	\centering
	\includegraphics[width=0.9\linewidth]{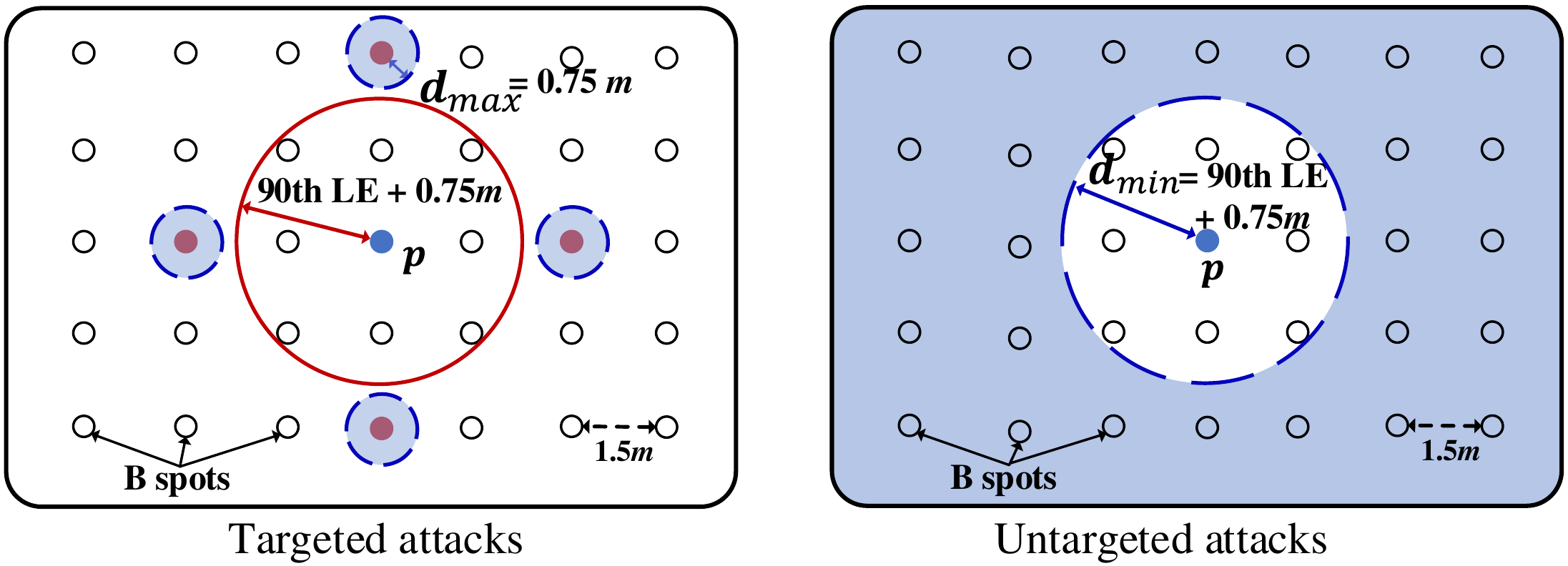}
	\caption{Illustration of our attack methodology adopted in offline experiments.}
	\label{fig:attackmethodology}
\end{figure}

\begin{table}
	\centering
	\caption{Performance of FooLoc in Offline Experiments.}\label{tab: performance offline}
	\begin{tabular}{|c|c|c|c|c|c|}
		\hline
		\multicolumn{6}{|c|}{Targeted attacks}                                                                                             \\ \hline
		\multicolumn{2}{|c|}{}             & \multicolumn{2}{c|}{Before}                   & \multicolumn{2}{c|}{After}                    \\ \hline
		\multicolumn{2}{|c|}{}             & $ \textit{DNN}_{A} $                     & $ \textit{DNN}_{B} $                    & $ \textit{DNN}_{A} $                     & $ \textit{DNN}_{B} $                     \\ \hline
		\multirow{2}{*}{\begin{tabular}[c]{@{}c@{}}LE to $ \mathbf{p} $\\ (Genuine spots)\end{tabular}}     & 50th     &            0.60 \text{m}           &          0.54 \text{m}             &            1.48 \text{m}            &             1.28 \text{m}           \\ \cline{2-6} 
		& 90th     &          1.85 \text{m}             &          1.93 \text{m}             &           2.61 \text{m}             &        2.51 \text{m}      \\ \hline
		\multirow{2}{*}{\begin{tabular}[c]{@{}c@{}}LE to $ \mathbf{q} $\\ (Targeted spots)\end{tabular}}     & 50th     &      1.59 \text{m}                    &          1.56 \text{m}               &          0.53 \text{m}                &             0.55 \text{m}           \\ \cline{2-6} 
		& 90th     &             3.08 \text{m}             &         2.93 \text{m}              &          1.42 \text{m}              &              1.38 \text{m}           \\ \hline
		\multicolumn{2}{|c|}{ASR} &         0.1\%               &            0.1\%             &          74.1\%             &             71.8\%          \\ \hline
		\multicolumn{2}{|c|}{PSR} &               -        &             -            &               -19.6 dB        &        -18.9 dB              \\ \hline
		\multicolumn{6}{|c|}{Untargeted attacks}                                                                                           \\ \hline
		\multicolumn{2}{|c|}{}             & \multicolumn{2}{c|}{Before}                   & \multicolumn{2}{c|}{After}                    \\ \hline
		\multicolumn{2}{|c|}{}             & $ \textit{DNN}_{A} $                     & $ \textit{DNN}_{B} $                    & $ \textit{DNN}_{A} $                    & $ \textit{DNN}_{B} $                     \\ \hline
		\multirow{2}{*}{LE to $ \mathbf{p} $}     & 50th     &         0.60 \text{m}               &            0.54 \text{m}            &           3.30 \text{m}              &          3.45 \text{m}              \\ \cline{2-6} 
		& 90th     &         1.85 \text{m}                &          1.93 \text{m}                &        5.55 \text{m}                 &       5.41 \text{m}                 \\ \hline
		\multicolumn{2}{|c|}{ASR} & 0.1\% & 0.0\% & 94.4\% & 92.4\% \\ \hline
		\multicolumn{2}{|c|}{PSR} & - & - & -19.0 dB & -19.5 dB \\ \hline
	\end{tabular}
\end{table}

\textbf{Attack Methodology.} We independently train $ \textit{DNN}_{A} $ and  $ \textit{DNN}_{B} $ on $ \mathcal{D}_{A} $, and optimize adversarial perturbations using the samples in $ \mathcal{D}_{B} $ according to Algorithm~1. Then, we apply the optimized perturbations on $ \mathcal{D}_{C} $ and feed the perturbed samples into $ \textit{DNN}_{A} $ and $ \textit{DNN}_{B} $, respectively, to perform both targeted and untargeted attacks. As depicted in Fig.~\ref{fig:attackmethodology}, for each B spot $ \mathbf{p} $ in targeted attacks, we choose the nearest B points that are outside a certain ball centering at $ \mathbf{p} $ as targeted spots. In particular, the ball radius equals to the sum of the 90th percentile LE of localization models and half of the spacing distance, i.e, $ 0.75 \; \text{m} $. In this way, we can have multiple targeted spots for one genuine spot $ \mathbf{p} $ and finally obtain 119 and 116 genuine-targeted spot pairs for $ \textit{DNN}_{A} $ and $ \textit{DNN}_{B} $, respectively. In addition, we configure $ d_{max} = 0.75\;\text{m} $ in targeted attacks. When performing untargeted attacks on $ \mathbf{p} $, we set $ d_{min} $ to be the sum of 90th percentile LE at $ \mathbf{p} $ of localization models and half of the spacing distance.     

\textbf{Experimental Results.} We first show the overall attack performance of FooLoc on $ \textit{DNN}_{A} $ and  $ \textit{DNN}_{B} $. For this purpose, we report all evaluation metrics in Table~\ref{tab: performance offline}. Before attacks, $ \textit{DNN}_{A} $ and $ \textit{DNN}_{B} $ obtain 50th LEs of 0.60~\text{m} and 0.54~\text{m}, respectively, which are comparable to other localization DNNs.
We can also observe that FooLoc has better performance in untargeted attacks in terms of LEs and ASRs. The reason is that FooLoc can search all directions pointing away from genuine spots in untargeted attacks, while having much fewer directions and more strict distance constraints to launch targeted attacks as shown in Fig.~\ref{fig:attackmethodology}. Despite that, in targeted attacks, $ \textit{DNN}_{A} $'s 90th percentile LE to genuine spots arises from 1.85~m to 2.61~m, while its 90th percentile LE to targeted spots decreases from 3.08~m to 1.42~m. Similar results can be found in $ \textit{DNN}_{B} $. Moreover, FooLoc achieves ASRs of 74.1\% and 71.8\% on $ \textit{DNN}_{A} $ and  $ \textit{DNN}_{B} $, respectively. The above observations suggest that FooLoc can effectively render victim models' predictions close to targeted spots. In untargeted attacks, FooLoc makes the 50th and 90th percentile LE of both models increase by over five and two times, respectively, implying that the two models' predictions are easily misled away from genuine spots. In addition, FooLoc obtains high ASRs of 94.4\% and 92.4\%, respectively, on $ \textit{DNN}_{A} $ and $ \textit{DNN}_{B} $ in untargeted attacks. It is worth noting that due to random noise and environmental dynamics, some collected Wi-Fi CSI samples may have already been predicted in targeted areas before adversarial attacks. However, such samples are only a very small portion of total testing samples, i.e., about 0.1\% as shown in Table~\ref{tab: performance offline}, which indicates the validity of targeted spot selection and acceptable distance error settings in our attack methodology. Furthermore, we also find that FooLoc has low PSRs of about -19~dB in both targeted and untargeted attacks. The result means that only small perturbations are introduced in original signals, which suggests the imperceptibility of our adversarial attacks. To sum up, the above results verify the effectiveness of FooLoc to deceive DL localization models. 

\begin{figure}[t]
	\centering
	\includegraphics[width=0.9\linewidth]{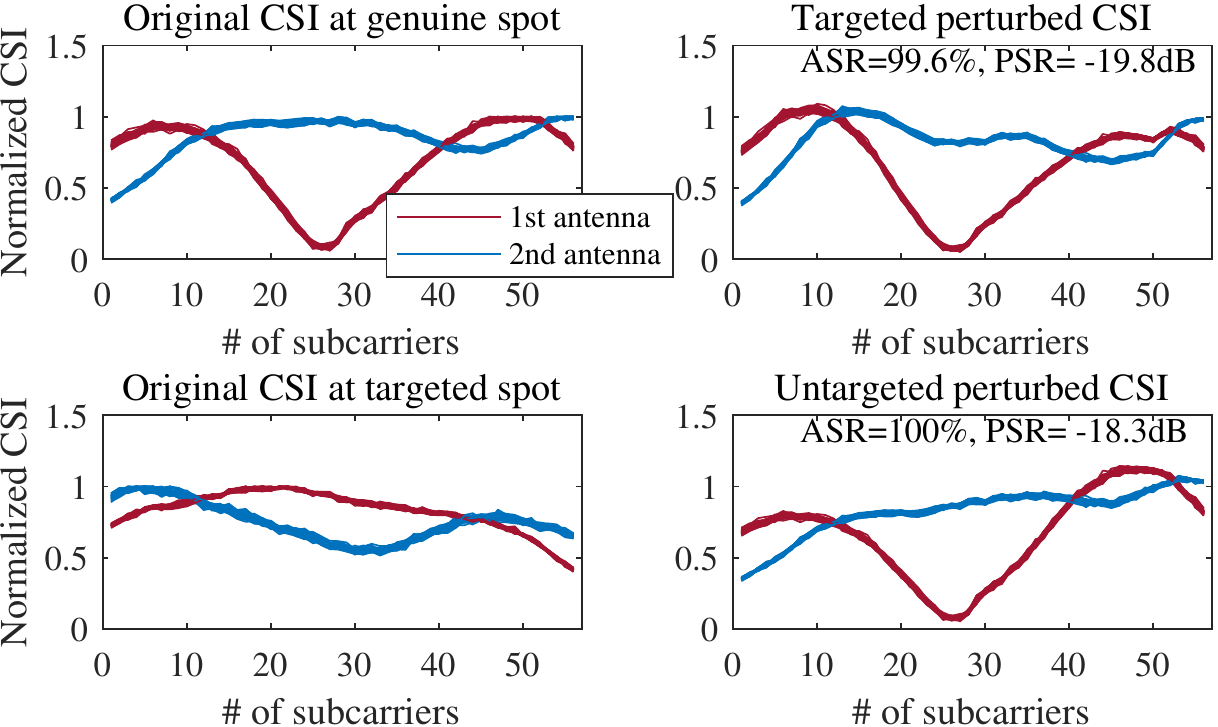}
	\caption{Illustration of original and perturbed signals under targeted and untargeted attacks in offline experiments.}
	\label{fig:offlinetargetedillustration}
\end{figure}

Next, we illustrate perturbed signals under targeted and untargeted attacks. Since FooLoc has similar attack performance on $ \textit{DNN}_{A} $ and $ \textit{DNN}_{B} $, we take perturbed signals of $ \textit{DNN}_{A} $ for illustration in Fig.~\ref{fig:offlinetargetedillustration}, where each subfigure depicts 50 CSI samples. As shown in Fig.~\ref{fig:offlinetargetedillustration}, we observe that under the same attack, the perturbed signals of two antennas share the same changing trends with respect to original ones. It is due to that our adversarial perturbations have multiplicative and repetitive impacts on original signals. Moreover, although the perturbed signals under two attacks are predicted to be far away from the genuine spot with high probabilities, they look very similar to original ones, which shows the usefulness of maximizing smoothness and limiting strength of adversarial weights in perturbation optimization. Furthermore, we can observe that targeted perturbed CSIs have more sudden changes and are less smoother when compared with untargeted perturbed CSIs. This is due to the fact that more changes are needed when FooLoc renders one sample to be estimated to come from a specified spot. Interestingly, we also find that targeted attacks have smaller perturbations on original signals. Though targeted perturbed signals show a very low similarity with original signals at the targeted spot, the corresponding predictions are less than 0.75~m from the targeted spot with a probability of 99.6\%. These observations suggest that localization DNNs are very vulnerable to our adversarial perturbations.

\begin{figure}[t]
	\centering
	\includegraphics[width=0.9\linewidth]{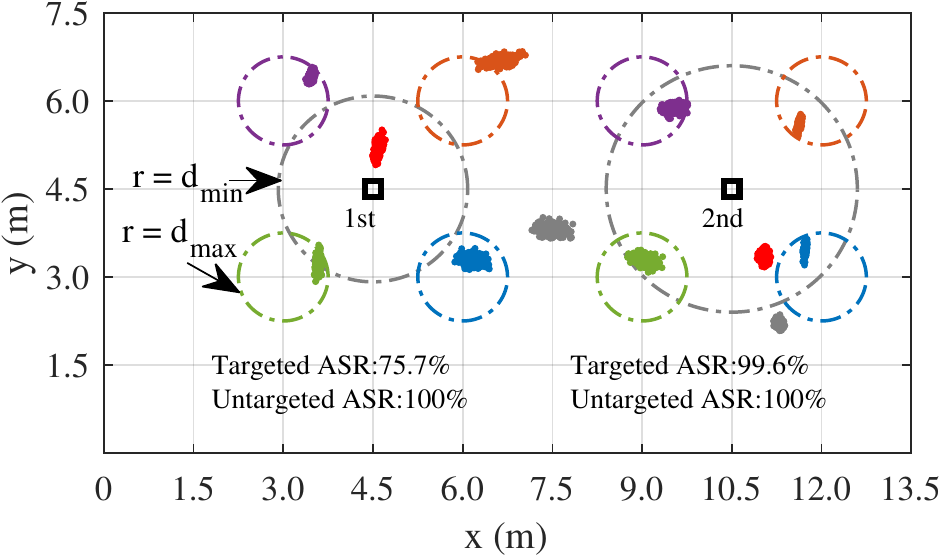}
	\caption{Illustration of adversarial attacks at two spots in the offline environment. The red dots are location predictions without perturbations. The gray dots are location predictions under untargeted attacks.}
	\label{fig:offline2dillustration}
\end{figure}

Then, we showcase FooLoc's targeted and untargeted attacks on $ \textit{DNN}_{A} $ at two B spots in the offline environment. To do this, we plot location predictions at two spots with and without adversarial attacks in the corresponding 2D Euclidean space in Fig.~\ref{fig:offline2dillustration}. In targeted attacks, the majority of CSI samples can be successfully perturbed into the neighboring area of targeted spots within a distance $ d_{max}=0.75~\text{m} $, even if these spots locate in different directions with respect to corresponding genuine spots. This observation verifies FooLoc's ability to render location predictions close to given targeted spots. In untargeted attacks, adversarial perturbations can make model predictions far away from genuine locations with a distance of more than $ d_{min} $. In addition, we can find that location predictions under untargeted attacks basically have a larger distance from genuine spots than that under targeted attacks. The above results illustrate the effectiveness of FooLoc to launch targeted and untargeted attacks on localization DNNs.

\begin{figure}
	\centering
	\includegraphics[width=0.9\linewidth]{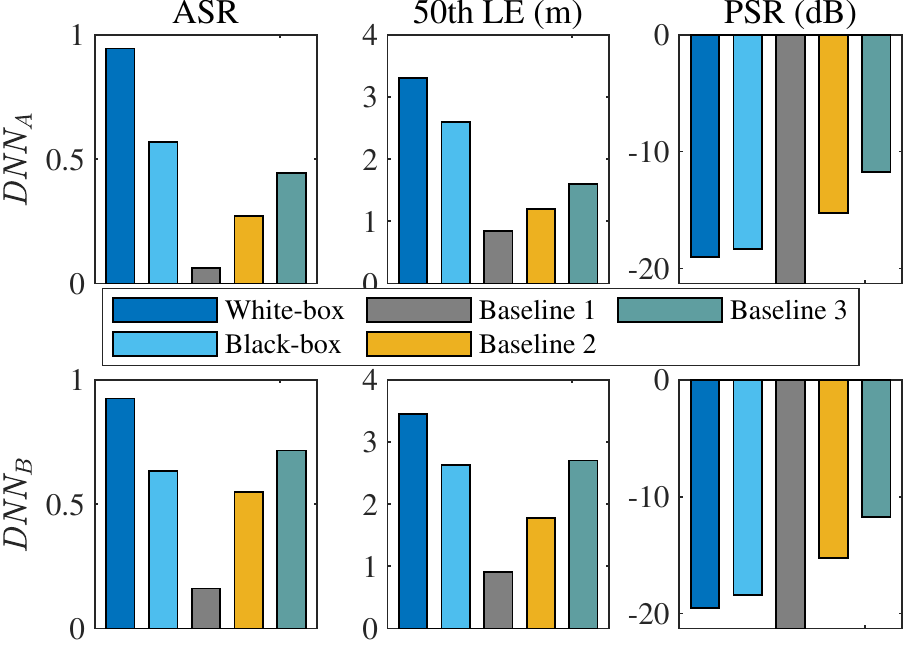}
	\caption{Performance of untargeted attacks under different conditions in offline experiments.}
	\label{fig:performanceblackbox}
\end{figure}

Furthermore, we show the feasibility of fooling black-box DL models over the air. In this case, the localization model used by the AP is unknown to the attacker. To simulate this situation, we first assume that $ \textit{DNN}_{A} $ is used by the AP. Then, we train $ \textit{DNN}_{A} $ using uplink CSI samples in the dataset $ \mathcal{D}_{A} $ as a victim model and optimize $ \textit{DNN}_{B} $ using the dataset $ \mathcal{D}_{B} $ as a substitute model. Next, we use the substitute model to generate untargeted adversarial perturbations with $ \mathcal{D}_{B} $ according to Algorithm~1. In this way, we can apply locally-generated perturbations on uplink CSI samples in $ \mathcal{D}_{C} $ to deceive unknown $ \textit{DNN}_{A} $. Similarly, we can attack $ \textit{DNN}_{B} $ if it is used by the AP using $ \textit{DNN}_{A} $ in a black-box manner. In this scenario, we also set three baseline models that leverage multiplicative perturbation weights randomly sampled from the interval $ \left( 1-\delta_{max}, 1+\delta_{max} \right) $. The baseline models, i.e., Baseline 1, Baseline 2 and Baseline 3, have different perturbation constraints $ \delta_{max} $ of 0.15, 0.3 and 0.45, respectively. During testing, we run each of them ten times and average all ASRs, 50th percentile LEs and PSRs as their final performance results.

As Fig.~\ref{fig:performanceblackbox} shows, FooLoc suffers performance degradation from white-box scenarios to black-box ones. These results are expected because the substitute models for perturbation generation in black-box attacks are different from targeted victim models, resulting in different adversarial weights. Moreover, when compared to other baseline models, Baseline 3 obtains the best performance in terms of ASRs and 50th LEs, while also having the highest PSRs. In addition, compared with Baseline 3, the black-box version of FooLoc achieves better performance on $ \textit{DNN}_{A} $ and comparable performance on $ \textit{DNN}_{B} $ with regard to ASRs and 50th LEs. However, it has much smaller PSRs on both two DNNs, suggesting that our adversarial attacks are more effective and stealthy than random perturbations. The above results indicate that FooLoc is capable of learning some shared adversarial weights that work well on different models due to the transferability of adversarial attacks~\cite{szegedy2013intriguing,goodfellow2014explaining}, showing the possibility of exploiting FooLoc to perform over-the-air adversarial attacks on black-box localization models.

\begin{figure}[t]
	\centering
	\includegraphics[width=0.9\linewidth]{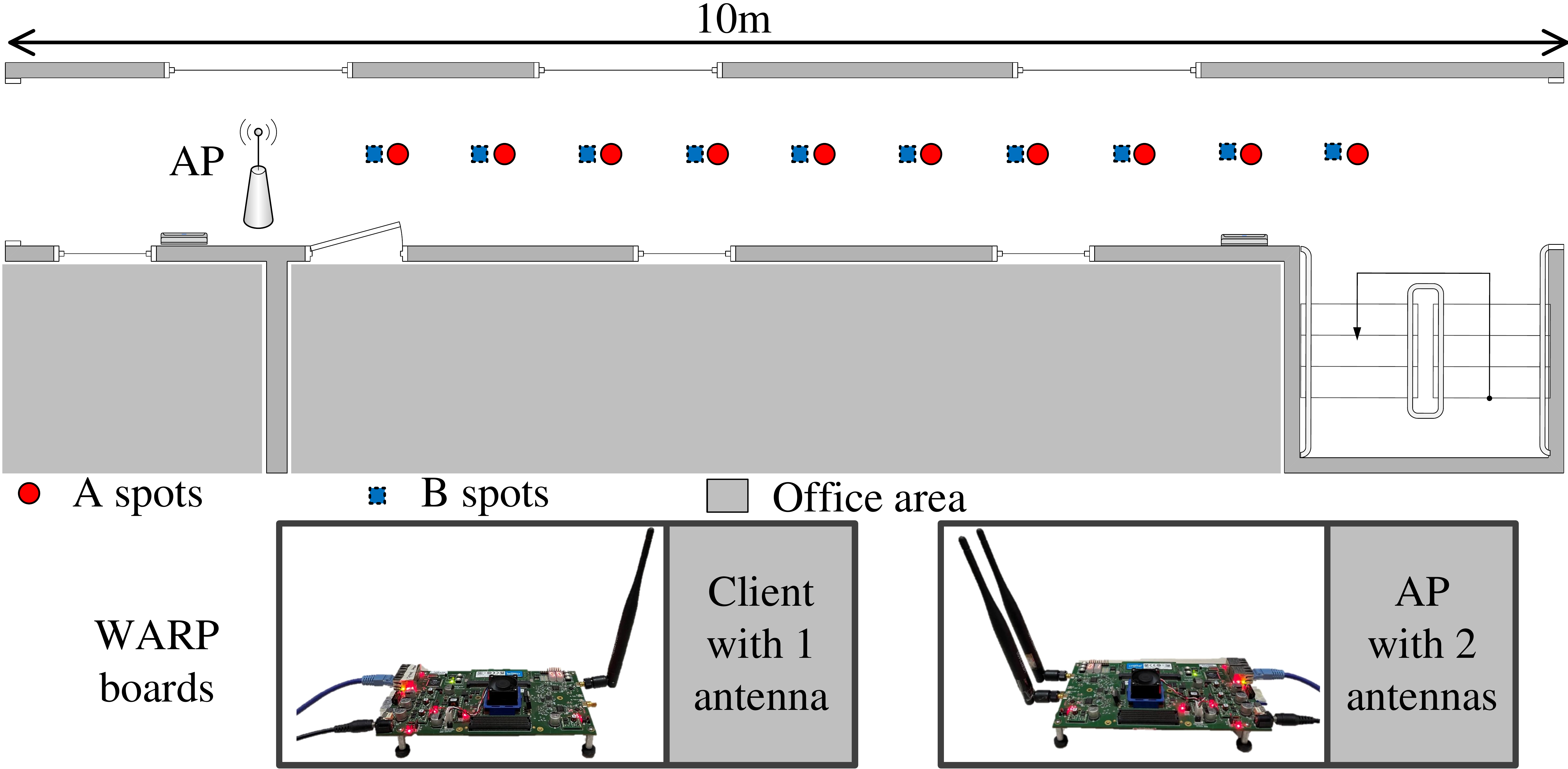}
	\caption{Floor plan of the experiment environment and experimental platform in online experiments.}
	\label{fig:floorplanonline}
\end{figure}

\subsection{Online Experiments}

In this subsection, we further examine the performance of FooLoc in online experiments. In this setting, we multiply adversarial weights with LTF signals, transmit perturbed signals to the AP over real wireless channels and record corresponding falsified uplink CSIs to attack localization models. 

\textbf{Implementation.} In online experiments, we implement FooLoc using the WARP wireless experimental platform~\cite{anand2010warplab} as shown in Fig.~\ref{fig:floorplanonline}. In particular, two WARP v3 boards are controlled by a Lenovo laptop via Ethernet cables to transfer control signals as well as their CSI measurements. One of the two boards is fixed at a certain location to act as an AP with two antennas, and the left board with one antenna works as a mobile client that communicates with the AP at the 5~GHz Wi-Fi band. Since WARP boards can provide channel estimates of 52 subcarriers, one CSI sample in online experiments has a size of $ 1 \times 2 \times 52 $.

\textbf{Data Collection.} We collect CSI measurements in a corridor environment as depicted in Fig.~\ref{fig:floorplanonline}. Specifically, we place the client at ten A spots and ten B spots in turn to record CSI measurements. First, we move the client among A spots, with a spacing distance of 0.6~m, and receive 1K uplink CSIs at each spot. In this way, we obtain a dataset $ \mathcal{D}_{E} $ containing 10K samples for training localization DNNs used by the AP. Then, by locating the client at B spots, we collect 1K downlink CSI samples at each location and have a dataset $ \mathcal{D}_{F} $ to generate adversarial perturbations. Note that there are stairs at one end of the corridor and people go downstairs and upstairs frequently. Thus, the collected CSI measurements are impacted by environmental noise and changes.

\textbf{Attack Methodology.} The attack strategy adopted in online experiments is similar to that in offline settings, but the only difference is that the attacker needs to send perturbed LTF signals over the air to deceive the victim AP. Specifically, we train $ \textit{DNN}_{A} $ and $ \textit{DNN}_{B} $, respectively, on the dataset $ \mathcal{D}_{E} $ and learn adversarial perturbations using $ \mathcal{D}_{F} $. Then, we multiply the locally-optimized perturbations on Wi-Fi LTF signals and transmit the perturbed signals from the client to the AP over the air. After the AP receives perturbed CSI measurements, we immediately feed them into $ \textit{DNN}_{A} $ and $ \textit{DNN}_{B} $, respectively, to perform location estimation. Moreover, we set $ d_{max} = 0.3~\text{m} $, i.e., the half of the spacing distance, and configure $ d_{min} $ to be the sum of the 90th percentile LE and $ d_{max} $. For a given B spot, the corresponding targeted spot is selected as a location that has a distance of 1.8~m from it.

\begin{figure}
	\centering
	\includegraphics[width=0.9\linewidth]{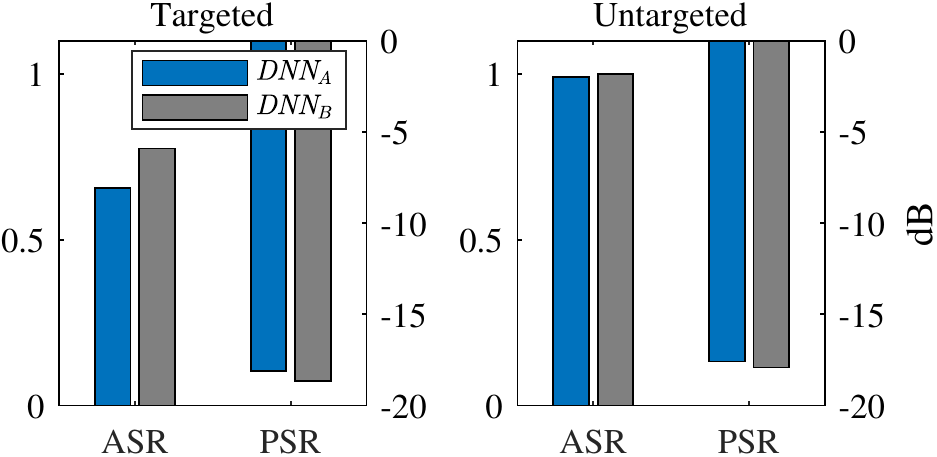}
	\caption{Attack performance of FooLoc in online experiments.}
	\label{fig:onlineoverallperformance}
\end{figure}

\textbf{Experimental Results.} We first report FooLoc's ASRs and PSRs in our online experiments. Since FooLoc's adversarial perturbations are learned from downlink CSI measurements, they would generally be affected by random environmental noise in uplink transmissions, resulting in performance degradation in terms of ASRs at the testing phase. As shown in Fig.~\ref{fig:onlineoverallperformance}, FooLoc achieves targeted ASRs of 65.7\% and 77.5\% on $ \textit{DNN}_{A} $ and $ \textit{DNN}_{B} $, respectively, which are comparable to that of FooLoc in offline experiments. In untargeted attacks, FooLoc obtains ASRs of above 99.0\% on two victim models, suggesting that FooLoc is still effective in this online setting. Moreover, our adversarial attacks have small perturbations on original signals and obtain mean PSRs of less than -17.5~dB in both targeted and untargeted scenarios. The above observations indicate that FooLoc is robust to environmental noise and has comparable performance in online experiments.  

Furthermore, different AP locations will impact FooLoc's performance. In general, the displacement of AP locations will produce different training sets of CSI fingerprints, which correspondingly changes the parameters of the localization model, thus resulting in different attack performance of our system. Roughly speaking, the higher localization accuracy the model achieves, the lower ASR FooLoc obtains. In our experiments, FooLoc achieves a targeted ASR of about 73\% and an untargeted ASR of about 93\% in the offline experiment, while obtaining a targeted ASR of about 71\% and an untargeted ASR of about 99\% in the online experiment. The above results show that FooLoc has similar attack performance in two different experimental settings.

\begin{figure}
	\centering
	\includegraphics[width=0.9\linewidth]{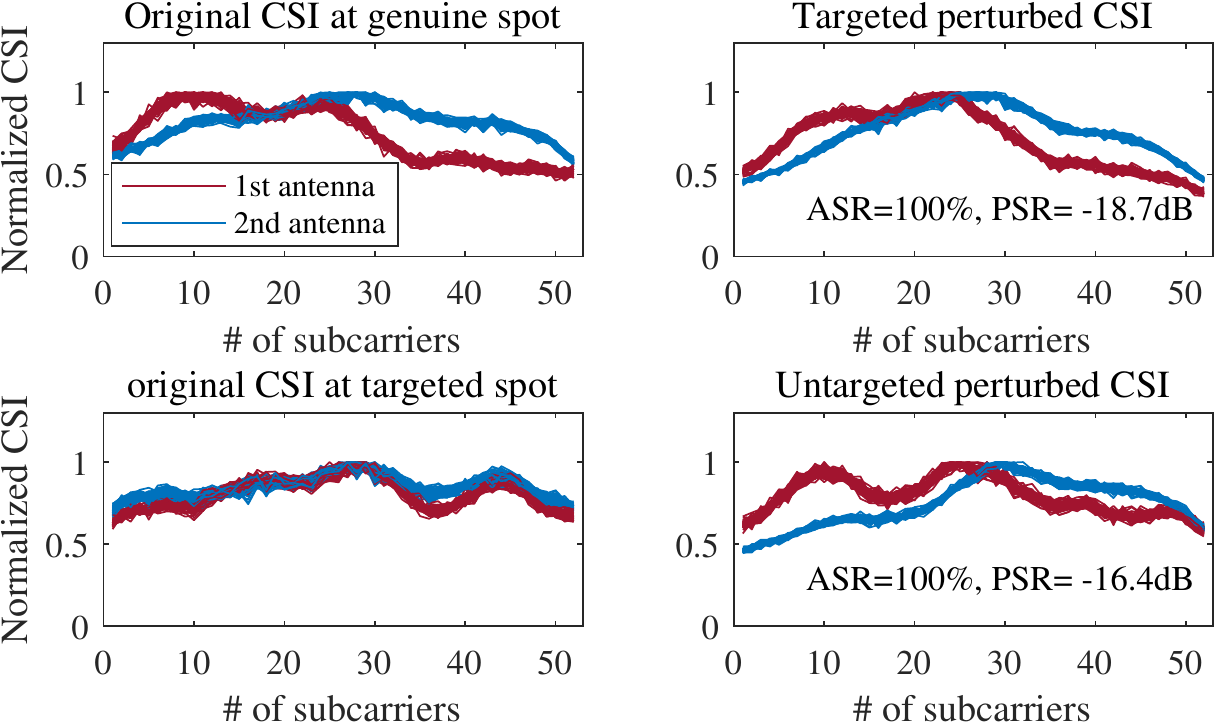}
	\caption{Illustration of original and perturbed signals under targeted and untargeted attacks in online experiments.}
	\label{fig:onlinesignalillustration}
\end{figure}

Next, we take a further step to show the imperceptibility of our adversarial perturbations. For this purpose, we record uplink CSI measurements at the AP with and without perturbations and depict corresponding signals for attacking $ \textit{DNN}_{A} $ in Fig.~\ref{fig:onlinesignalillustration}. As the figure shows, all perturbed CSI measurements look like original ones, i.e., keeping the main changing trends of original signals with slight differences. In addition, FooLoc can successfully generate adversarial signals with high ASRs and low PSRs. Although targeted perturbed CSIs are very different from original signals at the targeted spot, their predictions are less than 0.3~m from the targeted spot with a probability of 100\%. To sum up, our adversarial perturbations can effectively spoof DL localization models over realistic wireless channels.

\begin{figure}
	\centering
	\includegraphics[width=0.9\linewidth]{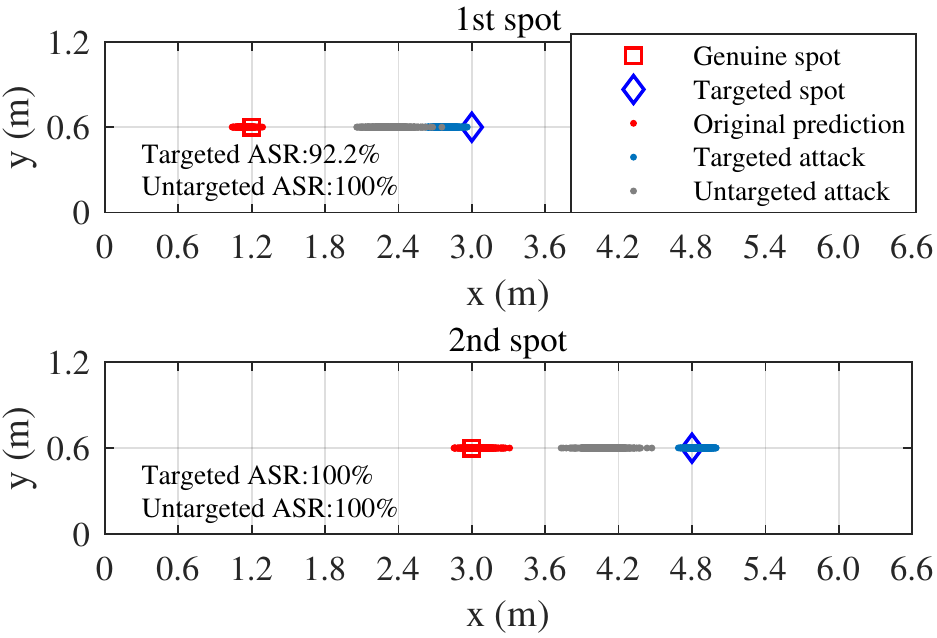}
	\caption{Illustration of adversarial attacks in the online environment. }
	\label{fig:online2dillustration}
\end{figure}

Then, we present location prediction results with and without adversarial attacks at two B spots in the online environment. To do this, we depict location predictions under adversarial attacks in the 2D Euclidean space in Fig.~\ref{fig:online2dillustration}. At the first spot, FooLoc can successfully render all location predictions in untargeted attacks far away from it with a distance of more than $ d_{min} $. At the same time, FooLoc makes location predictions in targeted attacks close to the targeted spot within a distance of 0.3~m with a high probability of 92.2\%. Similar observations can be also found in the second spot. The above results show the effectiveness of FooLoc to perform over-the-air targeted and untargeted adversarial attacks.

\section{Related Work} \label{sec:related work}

\textbf{Indoor Localization.} Recent years have witnessed the emerging needs of person or device locations in indoor environments, such as homes and office buildings~\cite{zhang2020peer, luo2021single, huang2020wifi}. Generally, indoor localization can be realized by exploiting various sensing modalities, among which Wi-Fi signals are one of the most promising ones thanks to their high ubiquity in indoor scenarios. Moreover, due to the huge success in the computer vision domain, various DNNs have been recently exploited for accurate Wi-Fi indoor localization~\cite{kotaru2015spotfi, xie2019md}. The stacked restricted Boltzmann machines~\cite{wang2015deepfi}, deep autoencoder~\cite{liu2018wifi} as well as residual networks~\cite{wang2020indoor} are proposed for indoor positioning, distance estimation, and so on. With the increasing usage of DNNs in indoor localization, it is thus of great importance to investigate the robustness of DL localization models to adversarial attacks.

\textbf{Adversarial Attacks.} Although deep neural networks have proven their success in many real-world applications, they are shown to be susceptible to minimal perturbations~\cite{szegedy2013intriguing, goodfellow2014explaining}. After that, various adversarial attacks are introduced in face recognition~\cite{sharif2016accessorize}, person detection~\cite{thys2019fooling}, optical flow estimation~\cite{ranjan2019attacking}, and so on. Recently, adversarial attacks are proposed on DNN based applications in wireless communications, such as radio signal classification~\cite{sadeghi2019adversarial}, waveform jamming and synthesis~\cite{restuccia2020generalized}. Moreover, the work~\cite{patil2021adversarial} exposes the threats of adversarial attacks on indoor localization and floor classification. However, this work uses additive perturbations, which can not tamper CSI measurements over realistic Wi-Fi channels. In our work, we propose multiplicative adversarial perturbations that can be exploited by adversary transceivers to perform adversarial attacks on localization DNNs over the air.

\textbf{Wireless Channel Manipulation.} Perturbations on wireless channels have also been investigated in the tasks of device authentication and device localization. 
Recently, researchers~\cite{cominelli2021ieee} propose a CSI randomization approach to distort location specific signatures for dealing with users' privacy concerns about locations. However, this approach lacks the capability of misleading location predictions close to specified spots, i.e., targeted attacks. In addition, the proposed random perturbations are not smooth, which will produce significant differences between perturbed CSI measurements and original ones, rendering them easy to be detected. However, FooLoc enables the attacker to launch both targeted and untargeted attacks, and our adversarial perturbations are smooth and minimal, making perturbed CSI signatures similar to the original ones. Moreover, the authors in~\cite{tung2016analog} propose analog man-in-the-middle attacks to mimic legitimate channel responses against link based device identification. The work~\cite{fang2014you} fools location distinction systems via creating virtual multipath signatures. These approaches trigger attacks via directly transforming genuine Wi-Fi CSI fingerprints to targeted ones, which is suitable for attacking single-antenna APs, which, however, are physically unrealizable in widely-used multi-antenna Wi-Fi systems due to the one-to-many relationship between the elements of one perturbation and one CSI measurement. In contrast, our attack takes this relationship into consideration and generates adversarial perturbations with a repetitive pattern, which characterizes the impact of over-the-air attacks on multi-antenna APs.

\section{Conclusion}\label{sec:conclusion}

This paper presents FooLoc, a novel system that launches over-the-air adversarial attacks on indoor localization DNNs. We observe that though the uplink CSI is unknown to FooLoc, its corresponding downlink one is obtainable and could be a reasonable substitute. Instead of using traditional additive perturbations, FooLoc exploits multiplicative perturbations with repetitive patterns, which are suitable for adversarial attacks over realistic wireless channels. FooLoc can efficiently craft imperceptible yet robust perturbations for triggering targeted and untargeted attacks against DL localization models. We implement our system using both commercial Wi-Fi APs and WARP v3 boards and extensively evaluate it in different indoor environments. The experimental results show that FooLoc achieves overall ASRs of about 70\% in targeted attacks and of above 90\% in untargeted attacks with small PSRs of about -18~dB. In addition, this paper reveals the bind spots of indoor localization DNNs using over-the-air adversarial attacks to call attention to appropriate countermeasures.


\bibliographystyle{IEEEtran}
\bibliography{IEEEabrv,./adversarial_attack}

\end{document}